\documentclass[pdflatex,sn-mathphys-num]{sn-jnl}

\usepackage{amsmath}        \usepackage{amsfonts}       \usepackage{bm}             \usepackage{mathcommand}    \usepackage{mathtools}      \usepackage{xparse}

\renewcommand*{\vec}[1]{{\bm{#1}}}

\newcommand*{\vb}{\vec{b}}

\newcommand*{\ve}{\vec{e}}

\newcommand*{\vg}{\vec{g}}

\newcommand*{\valpha}{\vec{\alpha}}

\newcommand*{\vnull}{\vec{0}}

\newcommand*{\mat}[1]{{\bm{#1}}}

\newcommand*{\mA}{\mat{A}}

\newcommand*{\mH}{\mat{H}}
\newcommand*{\mI}{\mat{I}}

\newcommand*{\mK}{\mat{K}}
\newcommand*{\mL}{\mat{L}}
\newcommand*{\mM}{\mat{M}}

\newcommand*{\set}[1]{{\mathbb{#1}}}

\newcommand*{\sR}{\set{R}}

         \newcommand*{\R}{\sR}

\newcommand*{\super}[2]{{#1^{#2}}}

\newcommand*{\transposesymbol}{\mathsf{T}}
\newcommand*{\hermitiansymbol}{\mathsf{H}}
\newcommand*{\adjointsymbol}{*}
\newcommand*{\inversesymbol}{-}

\newcommand*{\transpose}[1]{\super{#1}{\transposesymbol}}

\newcommand*{\hermitian}[1]{\super{#1}{\hermitiansymbol}}

\newcommand*{\inverse}[1]{\super{#1}{\inversesymbol 1}}

\newcommand*{\T}[1]{\transpose{#1}}
\renewmathcommand{\H}[1]{\hermitian{#1}}

\newcommand*{\inv}[1]{\inverse{#1}}

\newcommand*{\Tinv}[1]{\super{#1}{\inversesymbol\transposesymbol}}
\newcommand*{\adjinv}[1]{\super{#1}{\inversesymbol\adjointsymbol}}

\newcommand*{\invT}[1]{\Tinv}
\newcommand*{\invadj}[1]{\adjinv}
\newcommand*{\pinvT}[1]{\Tpseudoinv}
\newcommand*{\pinvadj}[1]{\adjpseudoinv}

\DeclarePairedDelimiter{\parens}{\lparen}{\rparen}
\DeclarePairedDelimiter{\bracks}{\lbrack}{\rbrack}
\DeclarePairedDelimiter{\braces}{\lbrace}{\rbrace}
\DeclarePairedDelimiter{\angles}{\langle}{\rangle}
\DeclarePairedDelimiter{\singlepipes}{|}{|}
\DeclarePairedDelimiter{\doublepipes}{\|}{\|}

\newcommand*{\dif}[2][1]{\mathrm{d}\ifstrequal{#1}{1}{}{^{#1}} #2}
\newcommand*{\pdif}[2][1]{\partial\ifstrequal{#1}{1}{}{^{#1}} #2}
\newcommand*{\diff}[2][1]{\mathrm{d} #2\ifstrequal{#1}{1}{}{^{#1}}}
\newcommand*{\pdiff}[2][1]{\partial #2\ifstrequal{#1}{1}{}{^{#1}}}

\NewDocumentCommand{\uniform}{s o m m}{
  \IfBooleanTF{#1}{
    \IfNoValueTF{#2}{
      \mathcal{U}\parens*{#3, #4}
    }{
      \mathcal{U}\parens*{#2 | #3, #4}
    }
  }{
    \IfNoValueTF{#2}{
      \mathcal{U}\parens{#3, #4}
    }{
      \mathcal{U}\parens{#2 | #3, #4}
    }
  }
}

\NewDocumentCommand{\normal}{s o m m}{
  \IfBooleanTF{#1}{
    \IfNoValueTF{#2}{
      \mathcal{N}\parens*{#3, #4}
    }{
      \mathcal{N}\parens*{#2 | #3, #4}
    }
  }{
    \IfNoValueTF{#2}{
      \mathcal{N}\parens{#3, #4}
    }{
      \mathcal{N}\parens{#2 | #3, #4}
    }
  }
}

\NewDocumentCommand{\dirac}{s m}{
  \IfBooleanTF{#1}
    {\delta\parens*{#2}}
    {\delta\parens{#2}}
}

\NewDocumentCommand{\GP}{s m m}{
  \IfBooleanTF{#1}
    {\mathcal{GP}\parens*{#2, #3}}
    {\mathcal{GP}\parens{#2, #3}}
}

\NewDocumentCommand{\GF}{s m m}{
  \IfBooleanTF{#1}
    {\mathcal{GF}\parens*{#2, #3}}
    {\mathcal{GF}\parens{#2, #3}}
}

\DeclareMathOperator{\variance}{Var}
\DeclareMathOperator{\covariance}{Cov}

\NewDocumentCommand{\probabilityof}{s o m}{
  \IfBooleanTF{#1}
    {p\parens*{#3\IfValueT{#2}{|#2}}}
    {p\parens{#3\IfValueT{#2}{|#2}}}
}

\NewDocumentCommand{\proposalof}{s m}{
  \IfBooleanTF{#1}
    {q\parens*{#2}}
    {q\parens{#2}}
}

\NewDocumentCommand{\expectationof}{s m}{
  \IfBooleanTF{#1}
    {\mathbb{E}\bracks*{#2}}
    {\mathbb{E}\bracks{#2}}
}

\NewDocumentCommand{\varianceof}{s m}{
  \IfBooleanTF{#1}
    {\variance\parens*{#2}}
    {\variance\parens{#2}}
}

\NewDocumentCommand{\covarianceof}{s m m}{
  \IfBooleanTF{#1}
    {\covariance\parens*{#2, #3}}
    {\covariance\parens{#2, #3}}
}

\DeclareMathOperator*{\diag}{diag}
\DeclareMathOperator*{\argmax}{arg\,max}
\DeclareMathOperator*{\argmin}{arg\,min}
\DeclareMathOperator*{\linspan}{span}
\DeclareMathOperator*{\rank}{rank}
\DeclareMathOperator*{\trace}{tr}

\NewDocumentCommand{\diagof}{s m}{
  \IfBooleanTF{#1}
    {\diag\parens*{#2}}
    {\diag\parens{#2}}
}

\NewDocumentCommand{\argmaxof}{s o m}{
  \IfBooleanTF{#1}
    {\argmax\IfValueT{#2}{_{#2}}\parens*{#3}}
    {\argmax\IfValueT{#2}{_{#2}}\parens{#3}}
}

\NewDocumentCommand{\argminof}{s o m}{
  \IfBooleanTF{#1}
    {\argmin\IfValueT{#2}{_{#2}}\parens*{#3}}
    {\argmin\IfValueT{#2}{_{#2}}\parens{#3}}
}

\NewDocumentCommand{\linspanof}{s m}{
  \IfBooleanTF{#1}
    {\linspan\parens*{#2}}
    {\linspan\parens{#2}}
}

\NewDocumentCommand{\rankof}{s m}{
  \IfBooleanTF{#1}
    {\rank\parens*{#2}}
    {\rank\parens{#2}}
}

\NewDocumentCommand{\traceof}{s m}{
  \IfBooleanTF{#1}
    {\trace\parens*{#2}}
    {\trace\parens{#2}}
}

\NewDocumentCommand{\norm}{s o m}{
  \IfBooleanTF{#1}
    {\doublepipes*{#3}\IfValueT{#2}{_{#2}}}
    {\doublepipes{#3}\IfValueT{#2}{_{#2}}}
}

\NewDocumentCommand{\innerproduct}{s o m m}{
  \IfBooleanTF{#1}
    {\angles*{#3, #4}\IfValueT{#2}{_{#2}}}
    {\angles{#3, #4}\IfValueT{#2}{_{#2}}}
}

\NewDocumentCommand{\collection}{s m m m}{
  \IfBooleanTF{#1}
    {\braces*{#2, #3, \dots, #4}}
    {\braces{#2, #3, \dots, #4}}
}

\newcommand*{\pos}{x}
\newcommand*{\vel}{v}
\newcommand*{\apos}{\xi}

\newcommand*{\Apos}{\Xi}

\newcommand*{\bpos}{\vec{\pos}}
\newcommand*{\bvel}{\vec{\vel}}
\newcommand*{\bapos}{\vec{\apos}}

\newcommand*{\cpos}{\mathcal{X}}
\newcommand*{\cvel}{\mathcal{V}}

\newcommand*{\transitionkernel}{q}

\newcommand{\eventrate}{\lambda}
\newcommand{\eventratei}{\eta}
\newcommand{\Eventrate}{\Lambda}
\newcommand{\affeventrate}{\tilde{\eventrate}}
\newcommand{\aeventrate}{\bar{\lambda}}
\newcommand{\aeventratei}{\bar{\eventratei}}
\newcommand{\affeventratei}{\tilde{\eventratei}}

\newcommand{\caeventrate}{\aeventrate^{\text{c}}}
\newcommand{\caeventratei}{\aeventratei^{\text{c}}}
\newcommand{\eventtime}{\tau}
\newcommand{\deltarate}{\gamma}
\newcommand{\decay}{\beta}
\newcommand{\refreshrate}{\lambda_{\text{ref}}}

\NewDocumentCommand{\transitionkernelof}{s o m}{
    \IfBooleanTF{#1}
    {\transitionkernel\parens*{\IfValueTF{#2}{#2}{\cdot}|#3}}
    {\transitionkernel\parens{\IfValueTF{#2}{{#2}|}{\cdot|}#3}}
}

\newcommand{\target}{\pi}
\newcommand{\potential}{\Psi}
\newcommand{\transformedpotential}{\tilde{\potential}}
\newcommand{\surrogate}{\bar{\potential}}
\newcommand{\laplace}{\surrogate_{\text{L}}}
\newcommand{\gp}{\surrogate_{\text{GP}}}
\newcommand{\gpkernel}{k}
\newcommand{\gpmean}{m}
\newcommand{\gpobsstd}{\sigma_{\text{n}}}
\newcommand{\gpobsnise}{\varepsilon}
\newcommand{\gpstd}{\sigma_{\text{f}}}
\newcommand{\gplengthscale}{\ell}
\newcommand{\gphyperparams}{\vartheta}
\newcommand{\bgphyperparams}{\vec{\vartheta}}
\newcommand{\gpkernelparam}{\gpkernel_\gphyperparams}
\newcommand{\gpoutput}{y}
\newcommand{\bgpoutput}{\vec{\gpoutput}}
\newcommand{\gpinput}{\bapos}
\newcommand{\gpdata}{\mathcal{D}}
\newcommand{\kernelmatrix}{\mK}

\NewDocumentCommand{\gpkernelparamof}{s m m}{
    \IfBooleanTF{#1}
    {\gpkernelparam\parens*{{#2}, {#3}}}
    {\gpkernelparam\parens{{#2}, {#3}}}
}

\newcommand{\disp}{u}
\newcommand{\vdisp}{\vec{\disp}}
\newcommand{\obs}{\tilde{\disp}}
\newcommand{\vobs}{\tilde{\vdisp}}
\newcommand{\coeff}{\theta}
\newcommand{\vcoeff}{\vec{\coeff}}
\newcommand{\pfield}{E}

\newcommand{\x}{x}
\newcommand{\method}{\mathcal{M}} 
\usepackage{xspace}

\makeatletter
\newcommand{\etal}{\@ifstar\etalstar\etalnostar}
\newcommand{\etc}{\@ifstar\etcstar\etcnostar}
\newcommand{\iid}{\@ifstar\iidstar\iidnostar}
\makeatother

\newcommand{\etalstar}{et~al.}
\newcommand{\etalnostar}{et~al.\@\xspace}
\newcommand{\ie}{i.e.\@\xspace}
\newcommand{\eg}{e.g.\@\xspace}
\newcommand{\etcstar}{etc.}
\newcommand{\etcnostar}{etc.\@\xspace}
\newcommand{\iidstar}{i.i.d.}
\newcommand{\iidnostar}{i.i.d.\@\xspace}

\newcommand{\ZZS}{Zig-Zag sampler}
\newcommand{\BPS}{Bouncy particle sampler}
\newcommand{\PDMP}{PDMP}
\newcommand{\PDMPs}{PDMPs}
\newcommand{\MCMC}{MCMC}
\newcommand{\RWM}{RWM}
\newcommand{\ESS}{ESS}
\newcommand{\HMC}{HMC}
\newcommand{\NUTS}{NUTS}
\newcommand{\PDE}{PDE}
\newcommand{\PDEs}{PDEs}
\newcommand{\MAP}{MAP}
\newcommand{\GPabbr}{GP}
\newcommand{\RMSE}{RMSE}
\newcommand{\WD}{WD}

\usepackage{graphicx}\usepackage{multirow}\usepackage{amssymb}\usepackage{amsthm}\usepackage{mathrsfs}\usepackage[title]{appendix}\usepackage{xcolor}\usepackage{textcomp}\usepackage{manyfoot}\usepackage{booktabs}\usepackage{algorithm}\usepackage{algorithmicx}\usepackage{algpseudocode}\usepackage{listings}

\usepackage{nicefrac}
\usepackage{microtype}
\usepackage{natbib}
\usepackage{url}
\usepackage{xfrac}
\usepackage{import}
\usepackage[group-separator={,}]{siunitx}
\usepackage{color}
\usepackage{subcaption}
\usepackage[colorinlistoftodos]{todonotes}
\usepackage{textcase}
\usepackage{dsfont}
\usepackage{placeins}

\usepackage[capitalise]{cleveref}

\newcommand{\bs}[1]{\boldsymbol{#1}}

\newcommand{\rowlabel}[1]{\hspace*{-0.5cm}
    \begin{minipage}[b]{0.04\linewidth}\centering
    \rotatebox{90}{\scriptsize#1}\end{minipage}}

\newcommand{\columnlabel}[1]{\makebox[0.32\textwidth][c]{\scriptsize#1}}

\begin{document}

\title[Article Title]{Piecewise Deterministic Markov Processes for Bayesian Inference of PDE Coefficients}

\author*[1]{\fnm{Leon} \sur{Riccius}}\email{l.f.riccius@tudelft.nl}
\author[1]{\fnm{Iuri B.C.M.} \sur{Rocha}}\email{i.rocha@tudelft.nl}
\author[2]{\fnm{Joris} \sur{Bierkens}}\email{joris.bierkens@tudelft.nl}
\author[2]{\fnm{Hanne} \sur{Kekkonen}}\email{h.n.kekkonen@tudelft.nl}
\author[1]{\fnm{Frans P.} \spfx{van der} \sur{Meer}}\email{f.p.vandermeer@tudelft.nl}

\affil[1]{\orgdiv{Faculty of Civil Engineering and Geosciences}, \orgname{Delft University of Technology}, \orgaddress{\street{Stevinweg 1}, \city{Delft}, \postcode{2628 CN}, \country{The Netherlands}}}
\affil[2]{\orgdiv{Delft Institute of Applied Mathematics}, \orgname{Delft University of Technology}, \orgaddress{\street{Mekelweg 4}, \city{Delft}, \postcode{2628 CD}, \country{The Netherlands}}}

\abstract{
    We develop a general framework for piecewise deterministic Markov process (PDMP) samplers that enables efficient Bayesian inference in non-linear inverse problems with expensive likelihoods.
    The key ingredient is a surrogate-assisted thinning scheme in which a surrogate model provides a proposal event rate and a robust correction mechanism enforces an upper bound on the true rate by dynamically adjusting an additive offset whenever violations are detected.
    This construction is agnostic to the choice of surrogate and PDMP, and we demonstrate it for the Zig-Zag sampler and the Bouncy particle sampler with constant, Laplace, and Gaussian process (GP) surrogates, including gradient-informed and adaptively refined GP variants.
    As a representative application, we consider Bayesian inference of a spatially varying Young’s modulus in a one-dimensional linear elasticity problem.
    Across dimensions, PDMP samplers equipped with GP-based surrogates achieve substantially higher accuracy and effective sample size per forward model evaluation than Random Walk Metropolis algorithm and the No-U-Turn sampler.
    The Bouncy particle sampler exhibits the most favorable overall efficiency and scaling, illustrating the potential of the proposed PDMP framework beyond this particular setting.
}


\maketitle

\section{Introduction}
\label{sec:intro}
Markov chain Monte Carlo (\MCMC) methods form a cornerstone of numerous scientific and engineering disciplines, including statistical physics, computational biology, and machine learning.
These methods are widely used for sampling from complex, typically unnormalized probability distributions that frequently arise in these contexts.
A particularly prominent application of \MCMC\ is Bayesian inference, where samples are drawn from posterior distributions to quantify uncertainties in model parameters based on observed data.

The majority of established \MCMC\ methods are built upon on the Random Walk Metropolis (\RWM) algorithm~\cite{Metropolis1953} and its generalization to non-symmetric proposals, the Metropolis-Hastings algorithm~\cite{Hastings1970}.
Despite their popularity due to simplicity and guaranteed asymptotic convergence, these algorithms suffer from poor sampling efficiency in high-dimensional parameter spaces due to their inherently diffusive nature~\cite{Roberts2001}.
To address these limitations, various advanced methods with improved sampling efficiency have emerged.
Notable examples include the first-order Metropolis-adjusted Langevin algorithm~\cite{Roberts2002}, Hamiltonian Monte Carlo (\HMC)\cite{Duane1987}, and their second-order manifold-aware extensions~\cite{Girolami2011}.
The No-U-Turn sampler (\NUTS)~\cite{Hoffman2014} further enhances \HMC\ by promoting large transitions without the need for manual tuning of trajectory lengths, and is widely adopted in probabilistic programming frameworks such as Stan~\cite{Carpenter2017} and PyMC~\cite{Salvatier2016}.
Empirical evaluations of these methods on practical engineering problems~\cite{Girolami2011,Chong2017,Goodman2010} confirm that their theoretically superior convergence properties observed on canonical test distributions~\cite{Gelman1997,Roberts1998} often carry over to realistic engineering applications.

All aforementioned methods share a common characteristic: reversibility.
Specifically, they satisfy the detailed balance condition, which ensures convergence to the target distribution but limits their sampling efficiency due to diffusive dynamics~\cite[ch 6.1]{Grimmett2016}.
Recently, there has been growing interest in irreversible sampling algorithms, motivated by theoretical results demonstrating their potential for faster convergence and lower asymptotic variance~\cite{Chen2013,Bierkens2016}.

Irreversible algorithms can be broadly categorized into two types.
One approach involves lifting reversible algorithms onto higher-dimensional state spaces~\cite{Ma2016}.
The second approach relies on piecewise deterministic Markov processes (\PDMPs).
PDMP-based samplers include the Zig-Zag sampler~\cite{Bierkens2017}, the Bouncy Particle sampler~\cite{Bouchard-Ct2018}, and the Boomerang sampler~\cite{Bierkens2020}.
These samplers evolve deterministically and are occasionally interrupted by stochastic events that abruptly alter their trajectories.
While these algorithms have favorable theoretical properties, such as good mixing, low autocorrelation, and error-free subsampling~\cite{Bierkens2019a}, their simulation is prohibitively expensive beyond a few simple examples of academic nature.
To date, \PDMP\ samplers have only seen limited application to practical problems, \eg linear Bayesian inverse problems~\cite{Ke2025}.

The \PDMP\ samplers define a non-homogeneous Poisson process.
In such a process, the time until next event is exponentially distributed with a rate that depends on the current state and the gradient of the log-density of interest.
A direct simulation of the next event time involves solving an integral equation, which has a closed form only for a few simple distributions, such as the Gaussian distribution with its quadratic log-density.
For more complex distributions, the next event time can be obtained via numerical integration of the event rate.
In~\cite{Bertazzi2022}, the \ZZS\ and \BPS\ are discretized with a forward Euler scheme, essentially defining a Markov chain on a grid of states.
Higher-order discretizations were considered in~\cite{Pagani2024}.
While numerical integration opens \PDMP\ sampling to a wide range of distributions, it is still not competitive in terms of computational cost in most cases.

The events can also be generated indirectly via Poisson thinning~\cite{Bierkens2019a}.
The thinning is applicable if an upper bound process is available, \ie a Poisson process with a rate strictly higher than the one of the process of interest.
A candidate event generated from this upper bound process is accepted with a probability equal to the ratio of the true rate and the upper bound.
Poisson thinning is only viable if events from the upper bound process can be generated efficiently.
In~\cite{Corbella2022}, a constant upper bound for a given time horizon is found by a polynomial fit to the switching rate of the \ZZS.
This process is repeated for each new state.

The direct inversion, as well as the thinning in~\cite{Corbella2022} can be summarized as purely local approaches for simulating \PDMPs, as they only consider the log-density and the event rate in the vicinity of the current state.
In this work, we propose a more global approach to \PDMP\ simulation, attempting to find an approximation of the event rate that can be used throughout the domain.
In theory, the event rate along the trajectory of the \PDMP\ is sufficient to generate the next event.
In practice, the information about the event rate in the rest of the domain can aid this event generation.

Our approach uses surrogate models to construct proposal rates for Poisson thinning for approximate \PDMP\ sampling.
A key feature is a correction mechanism that dynamically adjusts the surrogate whenever a violation of the upper bound condition is detected during simulation.
This correction mechanism renders the thinning procedure agnostic to the specific surrogate used:
the only requirement is that the surrogate induces a computable proposal rate.
This flexibility allows us to explore a range of surrogate families---from simple constant potentials or Laplace approximations to more expressive \GPabbr\ models---and to devise adaptive versions that incorporate information from the sampling process itself.

Our approach is tailored to unimodal, smooth posterior distributions defined on continuous parameter spaces.
Such distributions typically arise in the context of elliptic partial differential equations (\PDEs)---\eg heat conduction, electrostatics, or diffusion---and the inference of their coefficients.
We demonstrate and test our method on a simple yet illustrative example problem in linear elasticity, where we infer parameter fields from noisy displacement observations.
Our method is agnostic to the specific choice of \PDMP\ sampler.
We apply it to both the \ZZS\ and the \BPS\ and compare their performance.
Both samplers outperform a well-tuned \RWM\ baseline and the popular \NUTS\ across all considered metrics and settings, marking a step towards greater application of \PDMP\ sampling in Bayesian inference.

The remainder of this work is structured as follows.
In \cref{sec:pdmp}, we review the foundations of \PDMP\ sampling.
\Cref{sec:thinning_with_surrogates} introduces our surrogate-assisted \PDMP\ sampling framework and the surrogate model choices.
In \cref{sec:results}, we present numerical experiments demonstrating the performance of our method on a \PDE-governed Bayesian inverse problem.
Finally, \cref{sec:concusions} concludes the work with a summary and outlook on future research directions.

 \section{Piecewise-Deterministic Markov Process Samplers}\label{sec:pdmp}
\PDMPs\ are continuous-time Markov processes characterized by deterministic trajectories interrupted by random events.
We limit ourselves to \PDMPs\ designed for sampling from a target distribution \(\target\parens*{\bpos} \propto \exp\parens*{-\potential\parens*{\bpos}}\) on \(\cpos = \R^d\), where \(\potential:\R^d \to \R\) is a continuously differentiable potential.
A widely used subclass of \PDMP\ samplers, including the Zig-Zag and Bouncy particle samplers, is obtained by augmenting the position variable \(\bpos_t \in \cpos\) with a velocity variable \(\bvel_t \in \cvel = \R^d\)
Following~\cite{Fearnhead2018a}, a \PDMP\ is defined by these three components:
\begin{enumerate}
    \item \textit{The deterministic dynamics}: Between events, the position \(\bpos_t \in \cpos = \R^d\) and velocity \(\bvel_t \in \cvel = \R^d \) evolve deterministically according to
    \begin{equation}
        \frac{\dif{\bpos_t}}{\diff{t}} = \bs{\Phi}_{\bpos}\parens{\bpos_t, \bvel_t}, \quad
        \frac{\dif{\bvel_t}}{\diff{t}} = \bs{\Phi}_{\bvel}\parens{\bpos_t, \bvel_t},
    \end{equation}
    where \(\bs{\Phi}_{\bpos}\) and \(\bs{\Phi}_{\bvel}\) are vector fields governing the evolution.

    \item \textit{The event rate}: At switching times \(\eventtime\), the \PDMP\ undergoes a sudden transition.
    These events happen according to an inhomogeneous Poisson process with state-dependent, non-negative rate \(\eventrate\parens{\bpos_t, \bvel_t}\).

    \item \textit{The transition kernel}: When an event occurs, the \PDMP\ transitions from \((\bpos_t, \bvel_t)\) to \((\bpos_t^\prime, \bvel_t^\prime)\) according to a transition kernel \(\transitionkernelof*[\bpos_t^\prime, \bvel_t^\prime]{\bpos_t, \bvel_t}\).
\end{enumerate}
Different choices for \(\bs{\Phi}_{\bpos}\), \(\bs{\Phi}_{\bvel}\), \(\eventrate\), and \(\transitionkernel\), together with an initial condition \((\bpos_0, \bvel_0)\), fully define the \PDMP\@.

The Zig-Zag and Bouncy particle samplers feature piecewise constant velocities.
Between events, the dynamics are
\begin{equation}
    \frac{\dif{\bpos_t}}{\diff{t}} = \bvel_t, \hspace{2cm} \frac{\dif{\bvel_t}}{\diff{t}} = 0.
\end{equation}
The position \(\bpos_{t + s}\) at time \(t + s\) is straightforward in this case:
\begin{equation}
    \bpos_{t + s} = \bpos_t + s \cdot \bvel_t.\label{eq:pdmp-linear-dynamics}
\end{equation}
We will drop the time subscript for brevity in the following when there is no ambiguity.

These simple dynamics allow us to representgg the trajectory of the \PDMP\ entirely in terms of a skeleton \(\braces{\parens{t^k, \bpos^k, \bvel^k}}_{k\geq 0}\), which consists of event times \(t^k\) and the corresponding states \((\bpos^k, \bvel^k)\) immediately after each event.
Using \cref{eq:pdmp-linear-dynamics}, any position \(\bpos_t\) at time \(t \in \bracks{t^k, t^{k+1}}\) can then be linearly interpolated as \(\bpos_t = \bpos^k + \bvel^k \cdot \parens{t - t^k}\).

\subsection{The Zig-Zag Sampler}\label{subsec:the-zig-zag-sampler}
The velocity space of the \ZZS\ is discrete: \(\cvel = \{-1, +1\}^d\).
Events occur independently in each coordinate direction \(i\) with state-dependent intensity
\begin{equation}
    \eventratei_i(\bpos, \bvel) = \max \braces*{0, \vel_i \frac{\pdif{\Psi(\bpos)}}{\pdiff{\pos_i}} }.
\end{equation}
yielding a total event rate of
\begin{equation}
    \eventrate(\bpos, \bvel) = \sum_{i=1}^d \eventratei_i(\bpos, \bvel). \label{eq:zzs-event-rate}
\end{equation}
Events can be simulated from the total event rate~\eqref{eq:zzs-event-rate}.
The component to be flipped can then be selected with probability \(p_i\) proportional to its event rate
\begin{equation}
    p_i = \frac{\eventratei_i(\bpos, \bvel)}{\eventrate(\bpos, \bvel)}.
\end{equation}

Alternatively, the events can be simulated directly from the individual component rates.
The next switching time \(\eventtime\) is then given by the minimum of the switching times \(\eventtime_i\) across all components,
\begin{equation}
    \eventtime = \min_{i=1, \dots, d} \eventtime_i,
\end{equation}
and the process moves forward to \(\bpos' = \bpos + \eventtime \cdot \bvel\).

At the occurrence of an event, the transition kernel \(q\) flips the sign of a single component \(\vel_i\) of the velocity vector \(\bvel\).
The velocity right after the event \(\bvel'\) is given by
\begin{equation}
    \begin{aligned}
        \vel_i' &= -\vel_i,\\
        \vel_j' &= \vel_j \quad \forall j \neq i,
    \end{aligned}
\end{equation}
which yields the transition kernel $q$:
\begin{equation}
    \transitionkernelof*[\bvel']{\bpos', \bvel}
    = \dirac*{\bvel - 2 \ve_i \transpose{\ve_i} \bvel},
\end{equation}
where \(\ve_i\) is the \(i\)-th standard basis vector in \(\R^d\).
The position remains unchanged during the state update.
This construction yields a process with stationary distribution \(p(\bpos, \bvel) = p(\bpos) p(\bvel)\), where \(p(\bvel)\) is uniform over \(\{-1, +1\}^d\) and \(p(\bpos) = \target(\bpos)\).
The \ZZS\ is ergodic under mild conditions on the target distribution~\cite{Bierkens2019b}.

\begin{figure}
    \centering
    \subcaptionbox{\label{subfig:zzs_first_event}}{
        \includegraphics[width=0.4\textwidth]{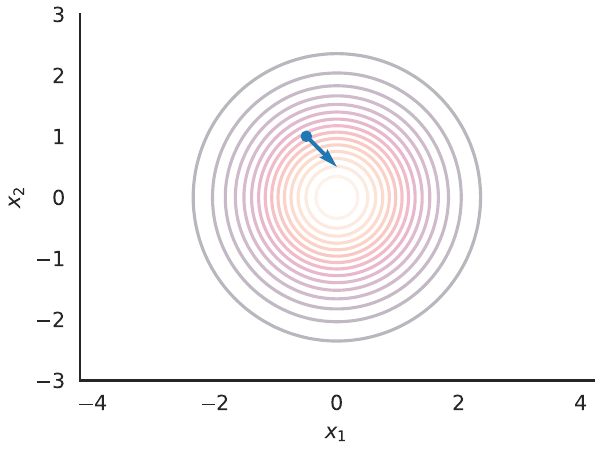}}\subcaptionbox{\label{subfig:zzs_first_rate}}{
        \includegraphics[width=0.4\textwidth]{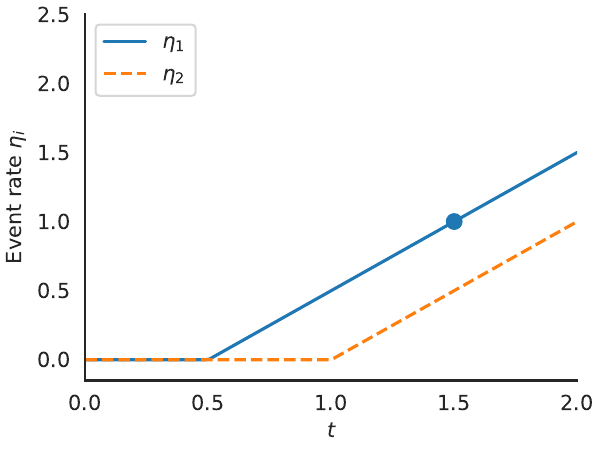}}\\
    \subcaptionbox{\label{subfig:zzs_second_event}}{
        \includegraphics[width=0.4\textwidth]{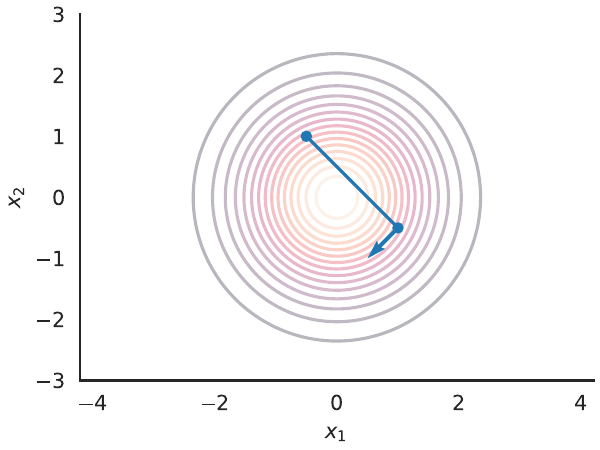}}\subcaptionbox{\label{subfig:zzs_second_rate}}{
        \includegraphics[width=0.4\textwidth]{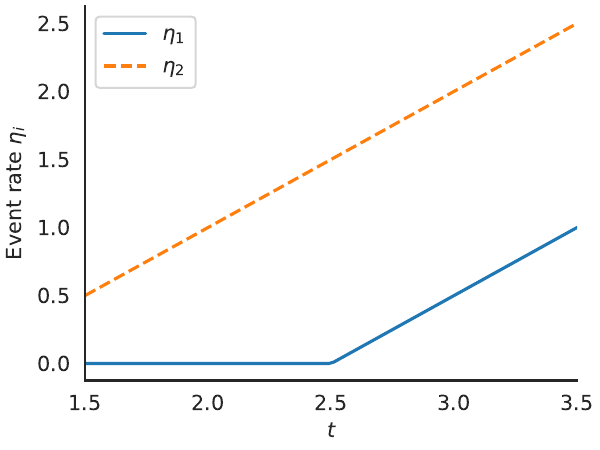}}\caption{
        Illustration of the \ZZS\ on a two-dimensional Gaussian.
        Figures (a) and (c) show the target density with the current position (dot) and velocity (arrow).
        Figures (b) and (d) show the corresponding event rates along each coordinate direction.
    }
    \label{fig:zzs-illustration}
\end{figure}

The movement of the \ZZS\ is guided by the gradient of the potential function \(\nabla \potential(\bpos)\).
When the sampler moves towards regions of higher density (\ie lower potential), the event rate is zero.
However, when the sampler moves towards the tails of \(\target(\bpos)\) (\ie higher potential), the event rate increases, leading to a higher chance of encountering an event.
Upon such an event, the respective velocity component flips, and the sampler is redirected back towards regions of higher density.
The position, velocity, and the resulting component-wise event rates of two consecutive events of the \ZZS\ on a two-dimensional Gaussian target are illustrated in \cref{fig:zzs-illustration}.
\begin{figure}
    \centering
    \subcaptionbox{\label{subfig:bps_first_event}}{
        \includegraphics[width=0.4\textwidth]{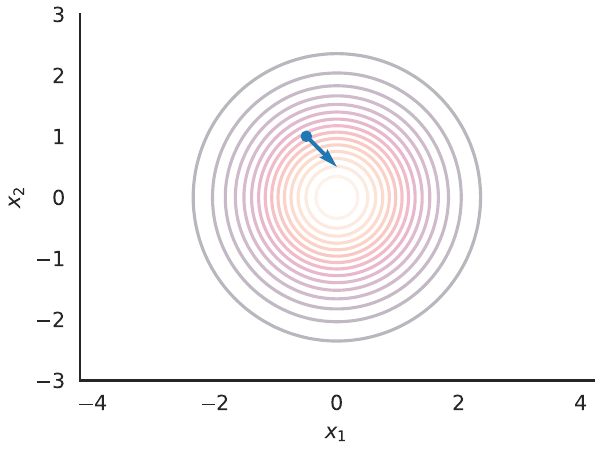}}\subcaptionbox{\label{subfig:bps_first_rate}}{
        \includegraphics[width=0.4\textwidth]{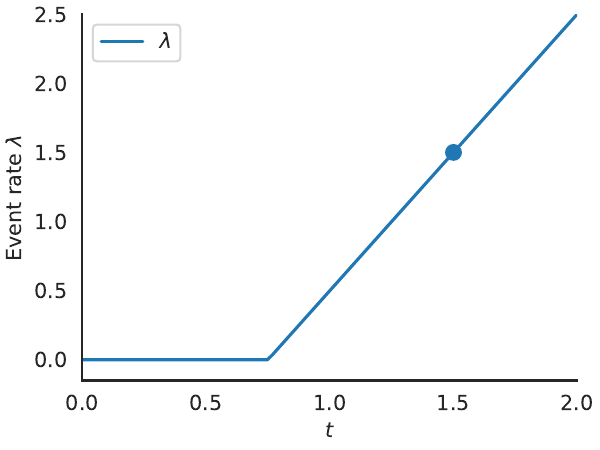}}\\
    \subcaptionbox{\label{subfig:bps_second_event}}{
        \includegraphics[width=0.4\textwidth]{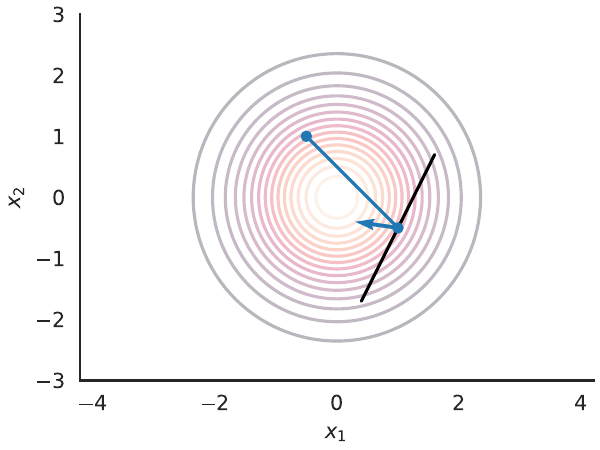}}\subcaptionbox{\label{subfig:bps_second_rate}}{
        \includegraphics[width=0.4\textwidth]{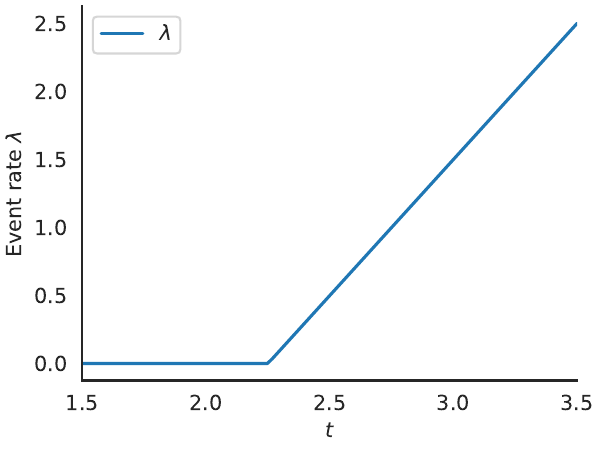}}\caption{
        Illustration of the \BPS\ on a two-dimensional Gaussian.
        Figures (a) and (c) show the target density with the current position (dot) and velocity (arrow).
        Figures (b) and (d) show the resulting event rate.
        Note that the event rate remains zero for longer, but has a steeper slope than in the \ZZS\@.
    }
    \label{fig:bps-illustration}
\end{figure}

\subsection{The Bouncy Particle Sampler}\label{subsec:the-bouncy-particle-sampler}
The \BPS\ employs a continuous velocity space \(\cvel = \R^d\).
Upon an event, the velocity is reflected against a hyperplane tangential to an isocontour of the target potential:
\begin{equation}
    \transitionkernelof*[\bvel']{\bpos', \bvel}
    = \dirac*{\bvel - 2\frac{\innerproduct*{\bvel}{\nabla \potential \parens*{\bpos'}}}{\norm*{\nabla \potential\parens*{\bpos'}}^2} \nabla \potential(\bpos')}.\label{eq:bps-reflection}
\end{equation}
This reflection flips the component in the direction of the gradient of the potential, effectively returning the particle towards regions of higher probability density.
Again, the position \(\bpos\) is unaffected by the event.
The reflection events occur with state-dependent intensity
\begin{equation}
    \eventrate\parens{\bpos, \bvel} = \max\braces*{0, \innerproduct*{\bvel}{\nabla \Psi(\bpos)}}.
\end{equation}
The bounce mechanism and event rate of the \BPS\ are illustrated on a two-dimensional Gaussian target in \cref{fig:bps-illustration}.

Because the original version of this algorithm~\cite{Peters2012} is not ergodic in certain scenarios, ~\cite{Bouchard-Ct2018} introduced random refreshments to ensure this sampler can visit the entire state space.
These random refreshment events, where the velocity is resampled from its stationary distribution \(p\parens{\bvel} = \normal{\vec{0}}{\mI}\), are exponentially distributed with rate \(\refreshrate > 0\).
If \(\refreshrate\) is too low, the sampler may struggle to reach certain regions of the state space.
If \(\refreshrate\) is too high, the sampler may exhibit diffusive behaviour, reducing its efficiency.
The optimal choice of \(\refreshrate\) is problem-dependent and often requires empirical tuning.

The \BPS\ can in theory find better paths through the state space on isotropic targets due to its continuous velocity space~\cite{Bierkens2025}.
However, one needs to carefully tune the refreshment rate \(\refreshrate\) to ensure ergodicity.
The \ZZS, on the other hand, does not require such tuning and runs out-of-the-box for a wide range of targets.

\subsection{Simulating Events with Poisson Thinning}\label{subsec:simulating-events}
Simulating event times from an inhomogeneous Poisson process with rate function \(\eventrate(\bpos_t, \bvel_t)\) can be challenging.
Two common methods for simulating these event times are the inverse method and the thinning algorithm (both described in~\cite{Lewis1979}).
The inverse method involves computing the integrated rate function
\begin{equation}
    \Eventrate(t) = \int_0^t \eventrate(\bpos_s, \bvel_s) \diff{s},
\end{equation}
and then sampling the next event time \(\eventtime\) by solving
\begin{equation}
    \Eventrate(\eventtime) = -\log u,\label{eq:inverse-method}
\end{equation}
where \( u \sim \uniform{0}{1} \).
This method requires that the integrated rate function \( \Eventrate(t) \) can be computed and inverted efficiently, which is generally not the case for non-linear \PDE-governed Bayesian inverse problems.

The thinning algorithm provides an alternative approach that does not require the inversion of the integrated rate function.
It involves simulating a candidate switching time \(\eventtime\) from a Poisson process with an upper bound \(\bar{\lambda}\) on the rate function \(\eventrate\).
The candidate switching time \(\eventtime\) is accepted with probability
\begin{equation}
    a(\eventtime) = \frac{\eventrate(\bpos_{t + \eventtime}, \bvel_{t + \eventtime})}{\bar{\lambda}(\bpos_{t + \eventtime}, \bvel_{t + \eventtime})}.
    \label{eq:thinning-acceptance-probability}
\end{equation}
If the candidate is rejected, we restart the event generation from time \(t + \eventtime\) without updating the velocity.

\begin{figure}
    \centering
    \subcaptionbox{Loose upper bound\label{subfig:thinning_loose_upper_bound}}{
        \includegraphics[width=0.5\textwidth]{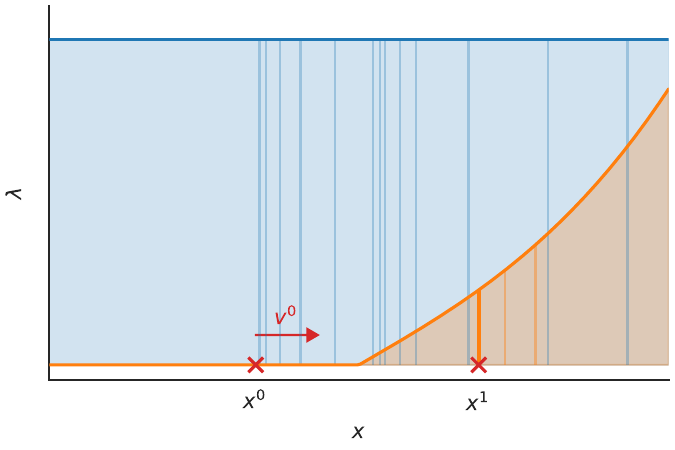}
    }\subcaptionbox{Tight upper bound\label{subfig:thinning_tight_upper_bound}}{
        \includegraphics[width=0.5\textwidth]{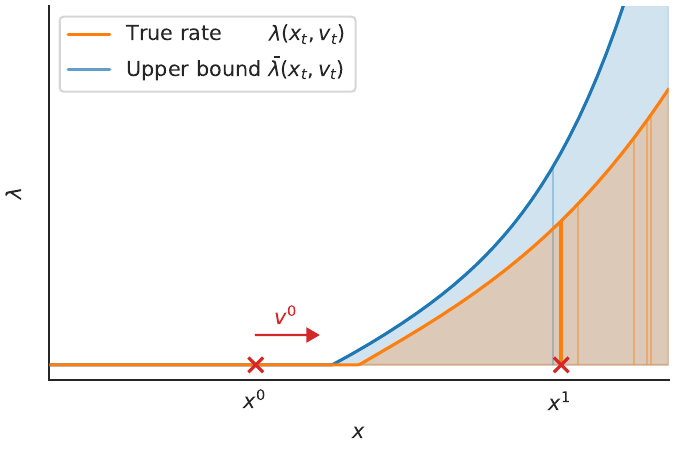}
    }\caption{
        Illustration of the thinning procedure for simulating event times.
        The vertical lines represent candidate event times drawn from a Poisson process with upper bound \(\bar{\lambda}\).
        Rejected events are shown in blue, while accepted events are shown in orange.
        Fewer rejections is better.
    }
    \label{fig:thinning}
\end{figure}

The thinning procedure is illustrated in \cref{fig:thinning} with a loose (left) and a tight (right) upper bound \(\bar{\lambda}\).
A tighter the upper bound, such as in \cref{subfig:thinning_tight_upper_bound}, yields a more efficient algorithm, as fewer candidate events will be rejected.
As \cref{eq:thinning-acceptance-probability} must be evaluated for each candidate event, having a smaller number of rejections is particularly attractive when each evaluation of the true event rate \(\eventrate\) is computationally expensive, \eg solving a \PDE\ for a realization of the parameter vector.

\subsection{Affine Transformation}\label{subsec:affine-transformation}
Both the \ZZS\ and the \BPS\ preserve their target distribution under affine reparameterizations of the state space~\cite{Bertazzi2022}.
More precisely, let
\begin{equation}
    \bapos = \inv{\mA}\, (\bpos - \vb),
\end{equation}
for some invertible matrix \(\mA \in \R^{d \times d}\) and vector \(\vb \in \R^d\).
Running the \ZZS\ or \BPS\ in the \(\bapos\)-coordinates with the transformed density
\begin{equation}
    \tilde{\pi}(\bapos) \propto \pi\parens*{\mA \, \bapos + \vb}
\end{equation}
and then mapping the resulting trajectory back via
\begin{equation}
    \bpos = \mA \, \bapos + \vb,
\end{equation}
yields a process in the original coordinates whose invariant distribution is \(\pi(\bpos)\).
This affine equivariance property may be exploited to improve sampling efficiency on anisotropic posterior distributions~\cite{Bierkens2025}.

In practice, we construct a local preconditioning transformation based on a Laplace approximation.
Let \(\bpos_{\text{MAP}}\) denote the maximum a posteriori estimate and let \(\mH=\nabla^2\potential(\bpos_{\text{MAP}})\) be the Hessian of the potential at that point.
We compute the Cholesky factorization \(\mH=\mL \, \T{\mL} \) and define the affine change of variables
\begin{equation}
    \bapos = \T{\mL}\, (\bpos - \bpos_{\text{MAP}}),
\end{equation}
or, equivalently, \(\bpos = \bpos_{\text{MAP}} + \Tinv{\mL} \, \bapos\), which centres the posterior at the MAP and locally whitens it with respect to the curvature of \(\potential\).
The transformed potential is then
\begin{equation}
    \transformedpotential \parens*{\bapos} = \potential \parens*{\bpos_{\text{MAP}} + \Tinv{\mL} \, \bapos}.
\end{equation}
In the general mapping \(\bpos = \mA\, \bapos + \vb\) above, this choice corresponds to \(\mA = \Tinv{\mL}\) and \(\vb = \bpos_{\text{MAP}}\).
We run the \ZZS\ or \BPS\ in the transformed space with density proportional to \(\exp(-\transformedpotential(\bapos))\), obtaining a trajectory \(\Apos_t\), and then map back via \(\bpos = \bpos_{\text{MAP}} + \Tinv{\mL} \, \bapos\).
 \section{Poisson Thinning with Surrogate Models}
\label{sec:thinning_with_surrogates}
Thinning requires an upper bound on the true PDMP event rate.
For the transformed potential \(\transformedpotential(\bapos)\), this transformed rate takes the form \(\affeventrate(\bapos, \bvel)\).
The main idea is to use a surrogate model \( \surrogate \parens{\bapos} \) to approximate the transformed potential \( \transformedpotential (\bapos) \).
We can then approximate the gradients of the potential \( \nabla \surrogate \parens{\bapos} \approx \nabla \transformedpotential \parens{\bapos} \)
and the process rate \( \aeventrate\parens{\bapos, \bvel} \approx \affeventrate\parens{\bapos, \bvel} \).
The cost of evaluating the surrogate is often orders of magnitudes lower than the cost of evaluating the true potential in \PDE-governed inverse problems.
Hence, we can use the direct inversion method and solve \cref{eq:inverse-method} numerically to produce event time candidates efficiently with the surrogate rate.
We then thin these candidates against the true event rate \( \affeventrate\parens{\bapos, \bvel} \) to obtain valid event times for the \PDMP\@.

In theory, the upper bound process must satisfy the global constraint
\begin{equation}
    \aeventrate\parens{\bapos, \bvel} \geq \affeventrate\parens{\bapos, \bvel},
    \label{eq:thinning_constraint}
\end{equation}
for all \( \parens{\bapos, \bvel} \), which guarantees the correctness of the thinning procedure.
In addition, we want this upper bound process to be as tight as possible for an efficient thinning procedure.

In practice, it is challenging to construct surrogate models that are both tight and satisfy the constraint in ~\cref{eq:thinning_constraint} globally.
We instead enforce the upper-bound condition only locally along the realized trajectory.
This is sufficient for correctness because event times in \PDMPs\ depend solely on the rate integrated along the current path.
A similar route is taken in~\cite{Corbella2022}, where local constant bounds for the event rate are constructed for a given time horizon.
We extend this idea by using global surrogate models that can adapt to the local behaviour of the potential function.
To ensure the correctness of the thinning procedure, we introduce a correction mechanism that adjusts the approximate rate \(\aeventrate\) whenever a violation of the upper bound condition is detected during the simulation.

Assuming that we violated the upper bound condition at some time \(t^k\) with position \( \bapos^k \) with velocity \( \bvel^k \), i.e. \( \aeventrate\parens{\bapos^k, \bvel^k} < \affeventrate\parens{\bapos^k, \bvel^k} \),
we can correct the surrogate model locally by adding an offset \( \deltarate \) to the event rate.
Formally, we define the corrected event rate as
\begin{equation}
    \aeventrate^\text{c} \parens{\bapos, \bvel} = \max\braces*{0, \innerproduct{\bvel}{\nabla \surrogate \parens{\bapos}} + \deltarate}, \label{eq:corrected_event_rate_bps}\\
\end{equation}
for the \BPS.
We initialize the offset to \( \deltarate = 0 \) at the start of the simulation.
In case of a violation of \cref{eq:thinning_constraint} at \( \parens{\bapos^k,\bvel^k} \), we increase the offset \(\deltarate\) as much as needed to satisfy the upper bound condition at the associated state:
\begin{equation}
    \deltarate =  \max \braces*{0, \affeventrate\parens{\bapos^k, \bvel^k} - \aeventrate\parens{\bapos^k, \bvel^k}}.\label{eq:offset_bps}
\end{equation}
The \ZZS\ is equipped with a vectorial offset \( \vec{\deltarate} \in \R^d \), with components updated similarly to the scalar offset in the \BPS, exploiting the additive structure of the \ZZS\ event rate:
\begin{align}
  \caeventratei_i \parens{\bapos, \bvel} &= \max \braces*{0, \vel_i \frac{\pdif{\surrogate(\bapos)}}{\pdif{\apos_i}} + \deltarate_i}, \label{eq:corrected_event_rate_zz} \\
  \caeventrate \parens{\bapos, \bvel} &= \sum_{i=1}^d \caeventratei_i \parens{\bapos, \bvel}, \\
  \deltarate_i &=  \max \braces*{0, \affeventratei_i\parens{\bapos^k, \bvel^k} - \caeventratei_i \parens{\bapos^k, \bvel^k}},
\end{align}
where \( \caeventratei_i \) is the component-wise proposal rate induced by the surrogate \( \surrogate\).

We revert the process to time \(t^{k-1}\), as the event that triggered the correction was invalid.
We then re-simulate the next event using either the corrected total event rate \( \caeventrate\) for the \BPS\ or its components \( \caeventratei_i\) for the \ZZS.
Note that we must also re-use the same outcome of the uniform random variable used for the candidate generation (see \cref{eq:inverse-method}).
Otherwise, we incur a bias towards sooner events, as these corrections occur more frequently when the PDMP travels further away from the mode of the target distribution.
Such long traverses are more likely for larger outcomes of the uniform random variable \( u\).

In its basic form, the offset \( \deltarate \) increases over time, making the thinning procedure less efficient.
To mitigate this issue, we implement a decay mechanism for the offset.
Denoting \( \deltarate^k \) the offset after the \(\parens{k - 1}\)-th event occurring at time \(t^{k-1}\),
we obtain the offset \( \deltarate^k \) that we use for the next interval \( [t^k, t^{k+1}) \) as
\begin{equation}
    \deltarate^k = \deltarate^{k-1} \cdot \exp \parens*{-\beta \parens*{t^k - t^{k-1}}},
    \label{eq:offset_decay}
\end{equation}
which means the offset decays exponentially over time with rate \( \beta > 0 \).
The discrete nature of the offset decay \cref{eq:offset_decay} means that the offset remains constant between events, and that we do not violate \cref{eq:thinning_constraint} when re-doing an event that triggered a correction.
We note that even with this correction, there is still no guarantee that the upper bound condition \cref{eq:thinning_constraint} holds for all \( \parens{\bapos, \bvel} \) along the trajectory, as we might miss fast oscillations of the rate between events.
Our method is therefore approximate in nature.
\cref{subsec:when_converge} holds a detailed discussion of the bias introduced by the approximate event generation scheme.

The complete algorithm for thinning with a corrected surrogate rate is detailed in \cref{alg:bps_thinning_surrogate} for the \BPS.
The counterpart for the \ZZS\ is given in \cref{alg:zzs_thinning_surrogate}.
The \ZZS\ version mainly differs in the way candidate event times are drawn and in the application of the correction mechanism.
We apply the same decay to each component of the vectorial offset \( \vec{\deltarate} \) as we did for the scalar offset in the \BPS in \cref{eq:offset_decay}.
Both \cref{alg:bps_thinning_surrogate} and \cref{alg:zzs_thinning_surrogate} return a skeleton of the trajectory, \ie the sequence of event times and corresponding states \(\braces{\parens{t^k, \bapos^k, \bvel^k}}_{k=0}^{K}\).
We typically set a final time \(T\) and run the algorithms until \(t_{K} \geq T\), as the \PDMP\ samplers converge in terms of simulated time, not simulated events~\cite{Bierkens2019b}.
The full trajectory can then be reconstructed easily by linear interpolation between event times.

\begin{algorithm}
  \caption{BPSWithSurrogate: scalar thinning with corrected surrogate rates}
  \label{alg:bps_thinning_surrogate}
  \begin{algorithmic}[1]
    \Require approximate proposal rate \(\aeventrate(\bapos,\bvel)\); true rate \(\affeventrate(\bapos,\bvel)\); initial state \(\parens{\bapos_0,\bvel_0}\) with \(\bvel_0\sim\normal{\vnull}{\mI}\); refresh rate \(\eventrate_\text{ref} > 0\); decay rate \(\beta\ge 0\); end time \(T\)
    \Ensure trajectory \(\braces{\parens{t^k,\bapos^k,\bvel^k}}_{k\ge 0}\)

    \State \(t\gets 0\); \(\parens{\bapos,\bvel}\gets\parens{\bapos_0,\bvel_0}\); \(k\gets 0\); \(\deltarate\gets 0\); \(\Delta_t \gets 0\)

    \While{\(t < T\)}
      \State \(\deltarate \gets \deltarate\, \exp\parens*{-\beta\, \Delta_t}\) \Comment{\textcolor{blue}{discrete decay}}

      \State \(\aeventrate^\text{c} \gets \max\braces{0, \innerproduct{\bvel}{\nabla \surrogate(\bapos)} + \deltarate}\)

      \State draw \(u_{\mathrm{ref}}\sim\uniform{0}{1}\); set \(\eventtime_{\mathrm{ref}} \gets -\log u_{\mathrm{ref}} / \eventrate_\text{ref}\) \Comment{\textcolor{blue}{refreshment clock}}
      \State draw \(u_\text{b} \sim \uniform{0}{1}\); find \(\eventtime_{\mathrm{b}}\ge 0\) solving \(\int_0^{\eventtime_{\mathrm{b}}} \aeventrate^\text{c}(\bapos+s\,\bvel,\bvel)\,\diff{s} = -\log u_\text{b}\) per \cref{eq:inverse-method}; set \(\bapos_p \gets \bapos + \eventtime_{\mathrm{b}}\,\bvel\); set event \(\gets\) bounce

      \While{true} \Comment{\textcolor{blue}{correction loop reusing \(u_\text{b}\)}}
      \If{\(\eventtime_{\mathrm{b}} > \eventtime_{\mathrm{ref}}\)}  \Comment{\textcolor{blue}{check refresh preemption}}
        \State event \(\gets\) refresh; \textbf{break}
      \EndIf
      \State \(\Delta_{\eventrate} \gets \affeventrate(\bapos_p,\bvel) - \aeventrate^c(\bapos_p, \bvel)\)
      \If{\(\Delta_{\eventrate} < 0\)}
        \State \textbf{break} \Comment{\textcolor{blue}{valid bounce candidate under bound}}
      \Else
        \State \(\deltarate \gets \deltarate + \Delta_{\deltarate}\); \(\aeventrate^c \gets \max\braces{0, \innerproduct{\bvel}{\nabla \surrogate(\bapos)} + \deltarate}\)
        \State recompute \(\eventtime_{\mathrm{b}}\) by solving \(\int_0^{\eventtime_{\mathrm{b}}} \aeventrate^c(\bapos+s\,\bvel,\bvel)\,\diff{s} = -\log u_\text{b}\)
        \State set \(\bapos_p \gets \bapos + \eventtime_{\mathrm{b}}\,\bvel\)
      \EndIf
      \EndWhile

      \If{event is refresh}
        \State \(t \gets t + \eventtime_{\mathrm{ref}}\); \(\bapos \gets \bapos + \eventtime_{\mathrm{ref}}\,\bvel\); \(\Delta_t \gets \eventtime_{\mathrm{ref}}\)
        \State resample \(\bvel \sim \normal{\vnull}{\mI}\) \Comment{\textcolor{blue}{refreshment}}
        \State \textbf{continue}
        \State \(k \gets k+1\); \(t^k \gets t\); \(\bapos^k \gets \bapos\); \(\bvel^k \gets \bvel\)

      \Else
        \State \(t \gets t + \eventtime_{\mathrm{b}}\); \(\bapos \gets \bapos_p\); \(\Delta_t = \eventtime_{\mathrm{b}}\)
        \If{\(u_\text{acc} \le \affeventrate(\bapos_p, \bvel)/\aeventrate^c(\bapos_p, \bvel)\)} \Comment{\textcolor{blue}{accept bounce}}
        \State draw \(u_\text{acc} \sim \uniform{0}{1}\)
        \State update \(\bvel\) by specular reflection per \cref{eq:bps-reflection}
        \State \(k \gets k+1\); \(t^k \gets t\); \(\bapos^k \gets \bapos\); \(\bvel^k \gets \bvel\);
        \EndIf
      \EndIf
    \EndWhile

    \State \textbf{return} trajectory \(\braces{\parens{t^k,\bapos^k,\bvel^k}}_{k\ge0}\)
  \end{algorithmic}
\end{algorithm}

\begin{algorithm}
  \caption{ZZSWithSurrogate: component-wise thinning with corrected surrogate rates}
  \label{alg:zzs_thinning_surrogate}
  \begin{algorithmic}[1]
    \Require approximate potential \(\surrogate(\bapos)\); true component rates \(\braces{\affeventratei_i(\bapos,\bvel)}_{i=1}^d\); initial state \(\parens{\bapos^0,\bvel^0}\) with \(\bvel_0\in\braces{-1,+1}^d\); decay rate \(\beta\ge 0\); end time \(T\)
    \Ensure trajectory \(\braces{\parens{t^k,\bapos^k,\bvel^k}}_{k\ge 0}\)

    \State \(t\gets 0\); \(\parens{\bapos,\bvel}\gets\parens{\bapos^0,\bvel^0}\); \(k\gets 0\); \(\Delta_t \gets 0\)
    \State initialize vector offset \(\vec{\deltarate}\gets \vnull\in\R^d\)

    \While{\(t < T\)}
      \State \(\vec{\deltarate} \gets \vec{\deltarate}\, \exp\parens*{-\beta\, \Delta_t}\)

      \State \(\caeventratei_i \gets \max \braces{0, \vel_i \pdif{\surrogate(\bapos)} / \pdif{\apos_i} + \deltarate_i}\) for \(i = 1,\dots,d\)

      \For{\(i=1,\ldots,d\)} \Comment{\textcolor{blue}{draw independent candidate times}}
      \State draw \(u_i\sim\uniform{0}{1}\)
      \State find \(\eventtime_i\ge 0\) solving \(\int_0^{\eventtime_i} \caeventratei_i(\bapos+s\,\bvel,\bvel)\,\diff{s} = -\log u_i\) per \cref{eq:inverse-method}
      \EndFor

      \While{true} \Comment{\textcolor{blue}{select earliest candidate; correct if bound violated}}
      \State \(i^* \gets \arg\min_i \eventtime_i\); \(\eventtime \gets \eventtime_{i^*}\); \(\bapos_p \gets \bapos + \eventtime\,\bvel\)
      \State \(\Delta_{\eventrate} \gets \affeventratei_{i^*}(\bapos_p, \bvel) - \caeventratei_{i^*}(\bapos_p, \bvel)\)
      \If {\(\Delta_{\eventrate} \le 0\)}
        \State \textbf{break} \Comment{\textcolor{blue}{valid candidate under bound}}
      \Else
        \State \(\deltarate_{i^*} \gets \deltarate_{i^*} + \Delta_{\eventrate}\); update \(\caeventratei_{i^*} \gets \max \braces{0, \vel_{i^*} \pdif{\surrogate(\bapos)} / \pdif{\apos_{i^*}} + \deltarate_{i^*}}\)
        \State find \(\eventtime_{i^*}\) solving \(\int_0^{\eventtime_{i^*}} \caeventratei_{i^*}(\bapos+s\,\bvel,\bvel)\,\diff{s} = -\log u_{i^*}\) \Comment{\textcolor{blue}{reuse \(u_{i^*}\)}}
      \EndIf
      \EndWhile

      \State \(t \gets t + \eventtime\); \(\bapos \gets \bapos_p\); \(\Delta_t \gets \eventtime\)
      \State draw \(u_\text{acc} \sim \uniform{0}{1}\)
      \If{\(u_\text{acc} \le \affeventratei_{i^*}(\bapos_p,\bvel)/\caeventratei_{i^*}(\bapos_p,\bvel)\)} \Comment{\textcolor{blue}{accept event}}
      \State flip component \(i^*\): \(\vel_i \gets - \vel_i\)
      \State \(k \gets k+1\); \(t^k \gets t\); \(\bapos^k \gets \bapos\); \(\bvel^k \gets \bvel\)
      \EndIf
    \EndWhile

    \State \textbf{return} trajectory \(\braces{\parens{t^k,\bapos^k,\bvel^k}}_{k\ge0}\)
  \end{algorithmic}
\end{algorithm}

\FloatBarrier

\subsection{Surrogate-Agnostic Design and Surrogate Families}\label{subsec:surrogate_agnostic}
A key advantage of the correction mechanism is that it guarantees correctness regardless of the surrogate model.
The only requirement is that the surrogate induces a computable proposal rate \(\aeventrate\parens{\bapos,\bvel}\); whenever this is not an upper bound, the local offset \(\deltarate\) in \cref{eq:corrected_event_rate_bps} restores correctness.
Consequently, the approach works even with extremely crude surrogates.
For example, taking a constant potential \( \surrogate(\bapos) \equiv c \) yields \( \nabla \surrogate \equiv \vnull \) and thus an initial proposal rate of zero.
In this case, we must initialize with a small non-zero offset (\eg \(\deltarate = 1\)) to generate candidate events from \(\aeventrate\); the first detected violation then raises the offset, after which the process proceeds correctly.
While such a baseline will be inefficient---offsets grow quickly, thinning acceptance is low, frequent evalutation of \(\affeventrate\)---it is valuable as a sanity check and lower-bound benchmark.

We consider the following surrogate families, ordered by increasing expressiveness and cost:
\begin{itemize}
  \item Constant potential (baseline): \(\surrogate(\bapos)\equiv c\).
  No training; correctness entirely via offsets.
  \item Laplace (trivial in transformed coordinates): after the affine transformation, the Laplace model reduces to the standard normal potential, \(\surrogate(\bapos) = \tfrac12\, \bapos^\top \bapos + c \), with gradient \(\bapos\) and Hessian \(\mI\), and arbitrary constant \(c\).
  This provides an inexpensive, stabilizing baseline around the mode.
  \item Laplace + \GPabbr\ residual: model the discrepancy \(r(\bapos) = \transformedpotential(\bapos) - \laplace(\bapos)\) with a \GPabbr pretrained on \(N_0\) realizations of \(r(\bapos)\).
  This combined model captures non-quadratic structure while preserving good local behaviour.
  \item Laplace + \GPabbr\ with gradient observations: enrich the \GPabbr\ training set with gradients of \(\transformedpotential\) where available, using derivative kernels.
  \item Adaptive \GPabbr\ (thinning-driven): start with \(N_0\) initial training points and expand the training set at milestone sizes \(N_0\,2^{n}\) for \(n=1,\ldots,5\), using data gathered from the thinning/simulation (e.g., locations of corrections or along accepted paths).
  This progressively tightens local rate proposals.
\end{itemize}
While we acknowledge the vast body of literature on more elaborate active learning schemes for \GPabbr s, we use the simple heuristic of the Adaptive \GPabbr\ model to demonstrate the potential of adaptive surrogates in this context.
Here, we include \textit{all} data gathered during simulation to expand the training set.
We therefore treat this model as an upper performance bound for active learning.

We follow a simple, consistent pipeline across surrogates:
first, we apply the affine transformation of \cref{subsec:affine-transformation} to whiten the state so that the Laplace approximation is standard normal in the transformed coordinate, as described in \cref{alg:affine}.
In these coordinates, the Laplace potential is trivial: \(\surrogate_{\mathrm{L}} = \tfrac12\,\bapos^\top\bapos+c\), with gradient \(\nabla \laplace(\bapos) = \bapos \) and Hessian \( \nabla^2 \laplace(\bapos) = \mI\), see \cref{alg:laplace}.
For all \GPabbr-based surrogates, we initialize by drawing \(N_0\) samples \(\{\bapos_i\}_{i=1}^{N_0} \sim \normal{\vnull}{\mI}\), evaluating \(\transformedpotential(\bapos_i)\), and training a \GPabbr\ on residuals \(r(\bapos_i)=\transformedpotential(\bapos_i)-\laplace\parens{\bapos_i}\), as in \cref{alg:gp}.
for the gradient-observing variant, we additionally include derivative observations \(\nabla r(\bapos_i)=\nabla\transformedpotential(\bapos_i)-\nabla\laplace\parens{\bapos_i}\).
For the adaptive, the \GPabbr\ hyperparameters are updated at each milestone, as long as the total number of data points does not exceed 1000,
as the computational costs grows cubically with the number of data points and further improvements are expected to be small in this data regime.

\begin{algorithm}
  \caption{ComputeAffineTransformation: affine transform to whiten the Laplace approximation}
  \label{alg:affine}
  \begin{algorithmic}[1]
    \Require target potential \(\potential(\bpos)\); initial guess \(\bpos^{(0)}\); tolerance \(\varepsilon\)
    \Ensure \MAP\ estimate \(\bpos_\text{MAP}\) and Cholesky factor \(\mL\) such that \(\mH = \mL\T{\mL} = \nabla^2\potential(\bpos_\text{MAP})\); transformed coordinate is \(\bapos = \T{\mL}(\bpos-\bpos_\text{MAP})\) as in \Cref{subsec:affine-transformation}
    \State Find \(\bpos_\text{MAP} \gets \argmin_{\bpos} \potential(\bpos)\) using a gradient-based optimizer (e.g., SciPy BFGS) starting at \(\bpos^{(0)}\); stop when \(\norm{\nabla \potential(\bpos)}<\varepsilon\)
    \State Compute analytic Hessian at the mode: \(\mH \gets \nabla^2 \potential(\bpos_\text{MAP})\)
    \State Cholesky factorization: \(\mH = \mL\T{\mL}\)
    \State \textbf{return} \(\bpos_\text{MAP},\mL\)
  \end{algorithmic}
\end{algorithm}

\begin{algorithm}
  \caption{BuildLaplaceSurrogate: quadratic surrogate in transformed coordinates}
  \label{alg:laplace}
  \begin{algorithmic}[1]
    \Require target potential \(\potential(\bpos)\); settings for \Cref{alg:affine}
    \Ensure Laplace surrogate \(\laplace(\bapos)\) in \(\bapos\)-space; gradient \(\nabla\laplace(\bapos)\); parameters \(\bpos_\text{MAP},\mL\)
    \State \(\bpos_\text{MAP},\mL \gets \Call{ComputeAffineTransformation}{\potential}\) \Comment{\textcolor{blue}{\Cref{alg:affine}}}
    \State Define the transformed potential implicitly by \(\tilde\potential(\bapos) \coloneqq \potential\bigl(\bpos_\text{MAP} + \Tinv{\mL}\bapos\bigr)\)
    \State Define \(\laplace(\bapos) \gets \tfrac12\, \T{\bapos}\bapos + \transformedpotential(\vec{0})\)
    \State Define \(\nabla\laplace(\bapos) \gets \bapos\)
    \State \textbf{return} \(\laplace,\nabla\laplace,\bpos_\text{MAP},\mL\)
  \end{algorithmic}
\end{algorithm}

\begin{algorithm}
  \caption{TrainGPSurrogate: \GPabbr\ residual on top of Laplace in transformed coordinates}
  \label{alg:gp}
  \begin{algorithmic}[1]
    \Require target potential \(\potential(\bpos)\); initial design size \(N_0\); kernel \(\gpkernel_\gphyperparams\) with init \(\bgphyperparams^{(0)}\); includeGrad \(\in\braces{\text{true,false}}\)
    \Ensure surrogate mean \(\gp(\bapos)=\laplace(\bapos)+\gpmean(\bapos)\) with gradient; parameters \(\bpos_\text{MAP},\mL\)
    \State \(\laplace,\nabla\laplace,\bpos_\text{MAP},\mL \gets \Call{BuildLaplaceSurrogate}{\potential}\) \Comment{\textcolor{blue}{\Cref{alg:laplace}}}
    \State Draw \(\braces{\bapos_i}_{i=1}^{N_0} \sim \normal{\vnull}{\mI}\) \Comment{\textcolor{blue}{training inputs in transformed space}}
    \For{\(i=1,\ldots,N_0\)}
      \State \(r_i \gets r(\bapos_i) \coloneqq \tilde\potential(\bapos_i) - \laplace(\bapos_i)\) \Comment{\textcolor{blue}{residual in transformed coords}}
      \If{includeGrad}
        \State \(\vg_{\xi,i} \gets \nabla_{\bapos} \tilde\potential(\bapos_i)\) \Comment{\textcolor{blue}{gradient in transformed coords via chain}}
        \State \(\vg_i \gets \vg_{\xi,i} - \nabla\laplace(\bapos_i)\) \Comment{\textcolor{blue}{residual gradient}}
      \EndIf
    \EndFor
    \State Assemble dataset \(\gpdata=\{(\bapos_i,r_i)\}\) (and gradient residuals if available)
    \State Fit \GPabbr\ prior \(f\sim\GP{0}{\gpkernel_\gphyperparams}\) by maximizing \(\mathcal{L}(\bgphyperparams)\)~\eqref{eq:gp_lml};
    \State Define \(\gpmean(\bapos) \gets \mathbb{E}[f(\bapos)\mid\gpdata,\hat{\bgphyperparams}]\) and \(\nabla\gpmean(\bapos)\) via~\eqref{eq:gp_grad_mean}
    \State Define composite surrogate \(\gp(\bapos)=\laplace(\bapos) + \gpmean(\bapos)\) with \(\nabla\gp(\bapos)=\nabla\laplace(\bapos)+\nabla\gpmean(\bapos)\)
    \State \textbf{return} \(\gp,\nabla\gp,\bpos_\text{MAP},\mL\)
  \end{algorithmic}
\end{algorithm}

\subsection{Gaussian Process Surrogate}\label{subsec:gp_surrogate}
We model a smooth scalar function on the transformed coordinates in the residual \(r(\bapos)=\transformedpotential(\bapos)-\laplace(\bapos)\) with a \GPabbr.
Let \( \braces{(\gpinput_i, \gpoutput_i)}_{i=1}^N\) denote training inputs and observations, where \(\gpoutput_i = f(\gpinput_i) + \epsilon_i\) with independent noise \(\gpobsnise_i \sim \normal{0}{\gpobsstd^2}\).
We place a constant-mean\footnote{
While the constant mean does not affect the gradients of the \GPabbr\ posterior mean, it can improve hyperparameter estimation and numerical stability \cite{Kennedy2001a}.
} \GPabbr\ prior \(f \sim \GP{m}{\gpkernelparam}\).
Let \(\gpinput, \gpinput' \in \R^d\) denote two input locations.
We use the anisotropic squared-exponential kernel
\begin{equation}
  \gpkernelparamof*{\gpinput}{\gpinput'}
  = \gpstd^2 \exp\parens*{ -\frac12 \sum_{j=1}^d \frac{(\xi_j - \xi'_j)^2}{\ell_j^2} },
  \qquad \bgphyperparams = \braces*{m, \gpstd^2, \gplengthscale_1,\ldots,\gplengthscale_d, \gpobsstd^2}.
  \label{eq:se_ard_kernel}
\end{equation}
The diagonal, dimension-specific length-scales \(\vec{\gplengthscale} = \braces{\gplengthscale_1,\dots,\gplengthscale_d}\) reflect anisotropy, which is necessary as the affine transformation in \cref{subsec:affine-transformation} only whitens the target in the vicinity of the posterior mode.
Together with the signal variance \(\gpstd^2\), observation noise \(\gpobsstd^2\), and mean parameter \(m\), this yields \(d+3\) hyperparameters, with \(d\) being the dimensionality of \(\bapos\).

We collect the kernel matrix \(\kernelmatrix \in \R^{N\times N}\) with entries \(\bracks{\kernelmatrix}_{ij} = \gpkernelparamof{\gpinput_i}{\gpinput_j}\) and let \(\kernelmatrix_\gpoutput = \kernelmatrix + \gpobsstd^2\mI\).
For a test location \(\gpinput_*\), we define the cross-covariance vector \(\kernelmatrix_* = \transpose{\bracks{\gpkernelparamof{\gpinput_*}{\gpinput_1},\ldots,\gpkernelparamof{\gpinput_*}{\gpinput_N}}}\) and the observations \(\bgpoutput = [\gpoutput_1,\ldots,\gpoutput_N]^\top\).
The \GPabbr\ posterior mean and variance are then:
\begin{align}
  \gpmean_*(\gpinput_*) &= m + \transpose{\kernelmatrix_*} \inverse{\kernelmatrix_\gpoutput} \parens*{\bgpoutput - m \mathbf{1}}, \\
  s_*^2(\gpinput_*) &= \gpkernelparamof{\gpinput_*}{\gpinput_*} - \transpose{\kernelmatrix_*} \inverse{\kernelmatrix_\gpoutput} \kernelmatrix_*,
  \label{eq:gp_posterior}
\end{align}
We evaluate \(\gpmean_*(\gpinput)\) as the surrogate mean and, when needed, its gradient to approximate the potential of the target distribution.

Gradients with respect to the input are available in closed form.
Writing \(\valpha = \inverse{\kernelmatrix_\gpoutput} \parens{\bgpoutput - m\mathbf{1}}\), the gradient of the predictive mean is
\begin{align}
  \nabla_{\gpinput} \gpmean_*(\gpinput) &= \sum_{i=1}^N \alpha_i \nabla_{\gpinput} \gpkernelparamof{\gpinput}{\gpinput_i} \label{eq:gp_grad_mean} \\
  \nabla_{\gpinput} \gpkernelparamof{\gpinput}{\gpinput_i} &= \gpkernelparamof{\gpinput}{\gpinput_i} \inverse{\mathrm{diag}(\vec{\gplengthscale})} \parens*{\gpinput_i - \gpinput},
\end{align}
noting that the constant mean parameter \(m\) does not depend on \(\gpinput\) and hence does not contribute to \(\nabla_{\gpinput} \gpmean_*(\gpinput)\).
If derivative observations are incorporated, we augment the \GPabbr\ with joint covariances obtained by differentiating the kernel: \(\mathrm{cov}(\partial_{p} f(\gpinput), f(\gpinput')) = \partial_{\gpinput_p} \gpkernelparamof{\gpinput}{\gpinput'}\) and
\begin{equation}
  \mathrm{cov} \parens*{\partial_{p} f(\gpinput),\, \partial_{q} f(\gpinput')}
  = \partial_{\gpinput_p}\partial_{\gpinput'_q} \gpkernelparamof{\gpinput}{\gpinput'}
  = \gpkernelparamof{\gpinput}{\gpinput'} \parens*{\frac{\delta_{pq}}{\ell_p^2} - \frac{(\gpinput_p-\gpinput'_p)(\gpinput_q-\gpinput'_q)}{\ell_p^2\,\ell_q^2}},
  \label{eq:gp_derivative_blocks}
\end{equation}
with \(\delta_{pq}\) the Kronecker delta and \(\partial_p\) the derivative w.r.t. the \(p\)-th input dimension.

Hyperparameter estimation proceeds by maximizing the marginal likelihood.
The log marginal likelihood for \(\vec{\gphyperparams}\) is
\begin{equation}
  \mathcal{L}(\vec{\gphyperparams})
  = \log \probabilityof[\gpinput_1,\ldots,\gpinput_N,\bgphyperparams]{\bgpoutput}
  = -\frac{1}{2} (\bgpoutput - m\mathbf{1})^\top \kernelmatrix_\gpoutput^{-1} (\bgpoutput - m\mathbf{1})\; - \frac{1}{2} \log\singlepipes{\kernelmatrix_\gpoutput}\; - \frac{N}{2}\log(2\pi),
  \label{eq:gp_lml}
\end{equation}
In practice, we compute \(\mathcal{L}\) and its gradients via a Cholesky factorization \(\kernelmatrix_\gpoutput = \mL \transpose{\mL}\), which yields \(\valpha = \Tinv{\mL}\inv{\mL}(\bgpoutput - m\mathbf{1})\) and \(\log\singlepipes{\kernelmatrix_\gpoutput} = 2\sum_i \log (\mL_{ii})\).
The gradient with respect to kernel and noise hyperparameters takes the standard form
\begin{equation}
  \frac{\pdif{\mathcal{L}}}{\pdif{\gphyperparams}_j}
  = \frac12\, \traceof*{\parens*{\valpha\transpose{\valpha} - \inv{\kernelmatrix_\gpoutput}}\, \frac{\pdif{\kernelmatrix_\gpoutput}}{\pdif{\gphyperparams_j}}},
  \label{eq:lml_grad_general}
\end{equation}
with
\begin{equation}
    \frac{\pdif{\kernelmatrix_\gpoutput}}{\pdif{\gpobsstd^2}} = \mI,
    \label{eq:lml_grad_noise}
\end{equation}
and kernel-specific derivatives \(\pdif{\bracks{\kernelmatrix_y}_{pq}}/\pdif{\gphyperparams_j} = \pdif{\gpkernelparamof{\gpinput_p}{\gpinput_q} / \pdif{\gphyperparams_j}}\) with
\begin{align}
  \frac{\pdif{\gpkernelparamof{\gpinput}{\gpinput'}}}{\pdif{\gpstd^2}} &= \frac{\gpkernelparamof{\gpinput}{\gpinput'}}{\gpstd^2}, \\
  \frac{\pdif{\gpkernelparamof{\gpinput}{\gpinput'}}}{\pdif{\ell_j}} &= \gpkernelparamof{\gpinput}{\gpinput'}\, \frac{\parens*{\xi_j-\xi'_j}^2}{\ell_j^{3}}.
  \label{eq:se_ard_kernel_derivatives}
\end{align}
The derivative with respect to the mean parameter is obtained from the quadratic term, \(\pdif{\mathcal{L}}/\pdif{m} = \transpose{\mathbf{1}}\,\kernelmatrix_\gpoutput^{-1}(\bgpoutput - m\mathbf{1})\).
 \section{Numerical Experiments}\label{sec:results}

\noindent
We begin by describing an experimental setup in \cref{subsec:test_case}, including a \PDE\ problem, performance metrics, and implementation details.
Focusing first on the \ZZS, we analyse when the proposed surrogate-based thinning method yields correct convergence in \cref{subsec:when_converge}.
\cref{subsec:surrogate_model_comparison} then compares the different surrogate models, followed by an analysis of the adaptive \GPabbr\ version in \cref{subsec:active_learning}.
Finally, \cref{subsec:zz_vs_bps_vs_nuts} benchmarks the Zig-Zag and Bouncy particle samplers against the No-U-Turn sampler~\cite{Hoffman2014}.

\subsection{Test Case}\label{subsec:test_case}
To demonstrate the proposed approach for \PDMP\ sampling, we introduce a scalable case study.
The problem is chosen to be computationally inexpensive, enabling direct comparison with long reference runs obtained using the Random-Walk Metropolis (\RWM) algorithm.
At the same time, the case study is motivated by its resemblance to practical engineering problems.
Its nonlinear parameter-to-observable map leads to a non-Gaussian posterior distribution, well-suited to test the capabilities of the \PDMP\ samplers and the surrogate-assisted framework.

\begin{figure}[H]
    \centering
    \subcaptionbox{d = 2\label{fig:bar_2d}}[0.49\textwidth]{{\scriptsize \def\svgwidth{\linewidth}
\begingroup%
  \makeatletter%
  \providecommand\color[2][]{%
    \errmessage{(Inkscape) Color is used for the text in Inkscape, but the package 'color.sty' is not loaded}%
    \renewcommand\color[2][]{}%
  }%
  \providecommand\transparent[1]{%
    \errmessage{(Inkscape) Transparency is used (non-zero) for the text in Inkscape, but the package 'transparent.sty' is not loaded}%
    \renewcommand\transparent[1]{}%
  }%
  \providecommand\rotatebox[2]{#2}%
  \newcommand*\fsize{\dimexpr\f@size pt\relax}%
  \newcommand*\lineheight[1]{\fontsize{\fsize}{#1\fsize}\selectfont}%
  \ifx\svgwidth\undefined%
    \setlength{\unitlength}{623.62204724bp}%
    \ifx\svgscale\undefined%
      \relax%
    \else%
      \setlength{\unitlength}{\unitlength * \real{\svgscale}}%
    \fi%
  \else%
    \setlength{\unitlength}{\svgwidth}%
  \fi%
  \global\let\svgwidth\undefined%
  \global\let\svgscale\undefined%
  \makeatother%
  \begin{picture}(1,0.20454545)%
    \lineheight{1}%
    \setlength\tabcolsep{0pt}%
    \put(0,0){\includegraphics[width=\unitlength,page=1]{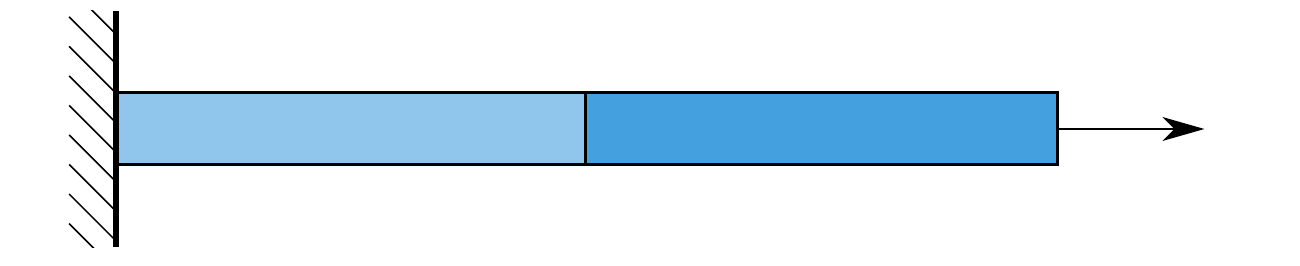}}%
    \put(0.65,0.16539781){\makebox(0,0)[lt]{\lineheight{1.25}\smash{\begin{tabular}[t]{l}$F(x=1) = \hat{F}$\end{tabular}}}}%
    \put(0.11333329,0.16539781){\makebox(0,0)[lt]{\lineheight{1.25}\smash{\begin{tabular}[t]{l}$u(x=0)=0$\end{tabular}}}}%
    \put(0.25293321,0.09195171){\color[rgb]{0,0,0}\makebox(0,0)[lt]{\lineheight{1.25}\smash{\begin{tabular}[t]{l}$E_1$\end{tabular}}}}%
    \put(0.61327501,0.09145785){\color[rgb]{0,0,0}\makebox(0,0)[lt]{\lineheight{1.25}\smash{\begin{tabular}[t]{l}$E_2$\end{tabular}}}}%
  \end{picture}%
\endgroup%
}}\hfill
    \subcaptionbox{d = 5\label{fig:bar_5d}}[0.49\textwidth]{{\scriptsize
            \def\svgwidth{\linewidth}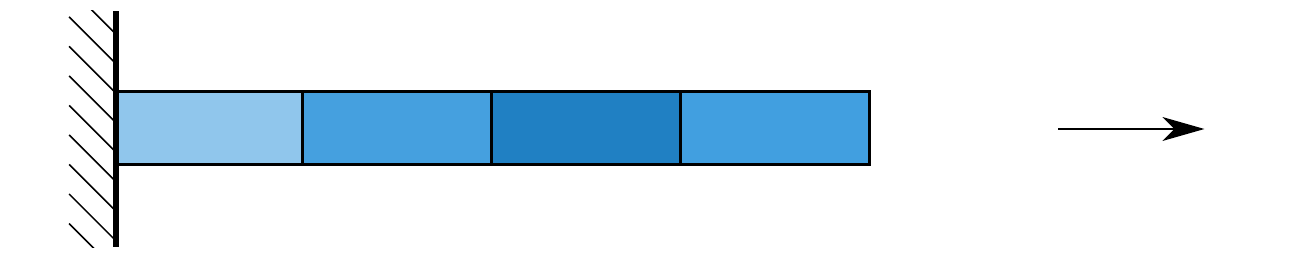}}

    \caption{Linear elastic bar problem with 2 and 5 dimensional discretization of Young's modulus field in (a) and (b), respectively.}
    \label{fig:1d_bar}
\end{figure}

Specifically, we consider the mechanical equilibrium of a one-dimensional linear elastic bar with a spatially varying and uncertain Young’s modulus, which is to be inferred from displacement measurements.
The bar with length L = 1 and constant cross-sectional area A = 1 is depicted in \cref{fig:1d_bar}.
The bar is fixed at the left end and subjected to a concentrated load \(\hat{F} = 1\) at the right end.
Assuming a linear elastic material, the displacement \(\disp(\x)\) along the bar is governed by the following boundary value problem:
\begin{align}
    \frac{\dif{}}{\dif{x}} \parens*{ \pfield(\x) \frac{\dif{\disp(\x)}}{\dif{\x}} } &= 0, \quad \x \in (0, 1), \\
    \disp(0) &= 0, \\
    \pfield(1) \frac{\dif{\disp(1)}}{\dif{\x}} &= 1,
\end{align}
where the Young's modulus \(\pfield(\x)\) depends on the position \(\x\).
We consider this parameter field \(\pfield(\x)\) to be uncertain.
In many situations however, some prior knowledge about the parameter field is available.

Here we adopt a smooth, spatially correlated prior field, which we model as a log-Gaussian random field, owing to the positivity constraint of the Young's modulus.
Such a field can be expressed as a Gaussian process:
\begin{align}
    \log \pfield(\x) &\sim \mathcal{GP}(\mu_{\pfield} (\x), k_{\pfield}(\x, \x')),\\
    k_{\pfield}(\x, \x') &= \sigma_{\pfield}^2 \exp \parens*{ -\frac{(\x - \x')^2}{2 l_{\pfield}^2} }.
\end{align}
Its three parameters are the mean function \(\mu_{\pfield}(\x)\), the signal variance \(\sigma_{\pfield}^2\), and the length scale \(l_{\pfield}\) of the squared exponential covariance function \(k_{\pfield}(\x, \x')\).
In practice, these parameters are estimated from data, \eg using maximum likelihood estimation, set based on expert knowledge, or treated as hyperparameters in a hierarchical Bayesian model.
For our testing purposes, we set \(\mu_{\pfield} = 1\), \(\sigma_{\pfield} = 1\) and \(l_{\pfield} = 0.3\), yielding a moderately varying field.

In theory, we would like to infer the continuous field \(\pfield(\x)\).
In practice, however, we need to discretize the field to represent it with a finite number of parameters \(\vcoeff\).
We discretize the field with piecewise-constant basis functions~\( \phi_i\)
\begin{align}
    \log \pfield(\x; \vcoeff) &= \sum_{i=1}^{d} \coeff_i \phi_i{\x},\\
    \phi_i(\x) &=
    \begin{cases}
        1, & \text{if } \dfrac{i-1}{d} \le \x < \dfrac{i}{d},\\
        0, & \text{elsewhere}.
    \end{cases}.
\end{align}
leading to a parameter vector \(\vcoeff \in \R^d\) of dimension \(d\).
We obtain the prior distribution of the parameters \(\probabilityof{\vcoeff} = \normal{\mu_{\vcoeff}}{\text{Cov}(\vcoeff)}\) by projecting the Gaussian process onto the piecewise-constant basis functions.
With \(M_{ij} = \langle \phi_i, \phi_j \rangle\), the covariance matrix of the parameters \(\vcoeff\) in the general case is given by
\begin{align}
    \mathrm{Cov}(\coeff_i, \coeff_j) &= \inv{\mM} \int_0^1 \int_0^1 \phi_i(\x) k_{\pfield}(\x, \x') \phi_j(\x') \, \diff{\x} \, \diff{\x'} \Tinv{\mM}.\\
\end{align}
Since the basis functions \(\phi_i\) are non-overlapping, and the basis is orthogonal with \(M_{ij} = \nicefrac{1}{d} \, \bs{I}\), the covariance simplifies to
\begin{align}
    \mathrm{Cov}(\coeff_i, \coeff_j) = d^2 \int_{(i-1)/d}^{i/d} \int_{(j-1)/d}^{j/d} k_{\pfield}(\x, \x') \, \diff{\x} \, \diff{\x'}.
\end{align}
With the constant mean function \(\mu_{\pfield}(\x) = 1\) of the underlying process, the mean of the coefficients \(\coeff_i\) is given by
\begin{align}
    \mathbb{E}[\coeff_i] &= d \int_{(i-1)/d}^{i/d} \pfield(\x) \, \diff{x} = 1.
\end{align}
Together with the feature vector \(\bs{\phi}(\x) = \T{\bracks{\phi_1(\x), \ldots, \phi_d(\x)}}\), the covariance of the discretized logarithm of the parameter field is given by
\begin{align}
    \mathrm{Cov}(\log \pfield(\x; \vcoeff), \log \pfield(\x'; \vcoeff)) &= \T{\bs{\phi}(\x)} \, \mathrm{Cov}(\vcoeff) \, \bs{\phi}(\x'),
\end{align}
which for \(d \to \infty\) converges to the original covariance function \(k_{\pfield}(\x, \x')\) of the Gaussian process.

We acknowledge that the piecewise-constant basis functions are not ideal for representing a smooth Gaussian process.
For most applications, other basis functions, such as Chebyshev polynomials or Karhunen-Loève basis functions, would be more suitable.
However, the piecewise-constant basis functions have a key advantage for our testing purposes:
they allow for an analytic solution of the forward problem, which reads
\begin{align}
    \disp(\x; \vcoeff) = \parens*{\sum_{i=1}^{\lfloor \x d \rfloor} \frac{\frac{1}{d}}{\exp\coeff_i}  +  \frac{\x - \frac{\lfloor \x d \rfloor}{d}}{\exp\coeff_{\lfloor \x d \rfloor + 1}} }, \quad \x \in [0, 1].
\end{align}
where \(\lfloor \cdot \rfloor\) denotes the floor function.
This solution is continuous and piecewise-linear, with kinks at the segment boundaries.
Most notably, the displacements are a non-linear function of the latent variables \(\vcoeff\), leading to a non-linear inverse problem.
Due to the analytic solution, we can evaluate the forward model and run the inference at negligible computational cost, allowing us to focus on the analysis of the sampling algorithms themselves.

First, we draw a sample from the prior \(\vcoeff^* \sim \probabilityof {\vcoeff}\) as the ground truth which we then try to infer from noisy displacement measurements \(\vobs\).
We generate these displacement measurements \(\obs_i\) by evaluating the forward model at \(m\) equidistant sensor locations \(\x_i = i/(m+1)\) for \(i=1,\ldots,m\) and adding independent Gaussian noise with standard deviation \(\sigma_{\mathrm{obs}} = 0.025\).
We therefore adopt a Gaussian likelihood model
\begin{align}
    p(\vobs | \vcoeff) = \mathcal{N}(\vobs | \vdisp(\vcoeff), \sigma_{\mathrm{obs}}^2 \, \bs{I}),
\end{align}
where \(\vobs = \T{\bracks{\obs_1, \ldots, \obs_m}}\) are the noisy measurements, and \(\vdisp(\vcoeff) = \T{\bracks*{\disp(x_1; \vcoeff), \ldots, \disp(\x_m; \vcoeff)}}\) are the model predictions at the sensor locations.
Following~\cite{Riccius2026}, we set the number of sensors to \(m = \lfloor \nicefrac{3}{4} d \rfloor\).
Fixing the observation-to-parameters ratio in this manner helps to maintain a consistent level of problem difficulty across different parameter dimensions \(d\).

\begin{figure}
    \centering
    \subcaptionbox{\label{fig:2d_no_affine}}[0.4\textwidth]{\includegraphics[width=\linewidth]{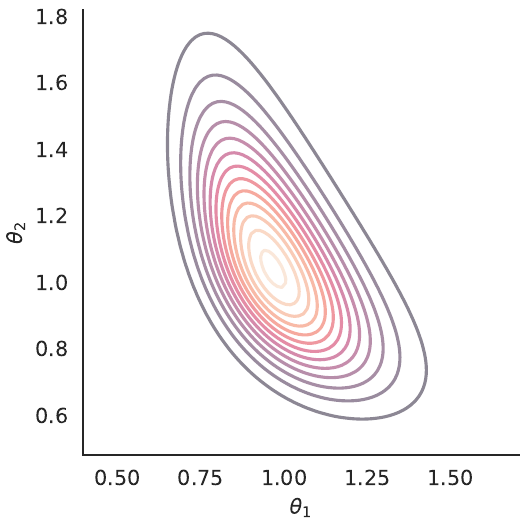}}\subcaptionbox{\label{fig:2d_affine}}[0.4\textwidth]{\includegraphics[width=\linewidth]{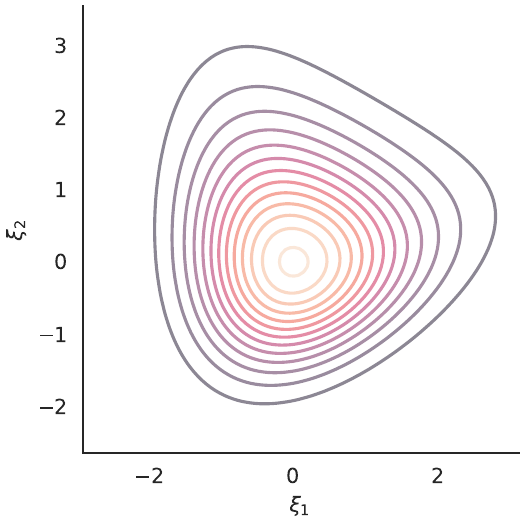}}\caption{Posterior distribution of the two-dimensional problem in the original (a) and the affine space (b).}
    \label{fig:posterior}
\end{figure}

Finally, we summarize the Bayesian inference problem as
\begin{align}
    \probabilityof[\vobs]{\vcoeff} &\propto \probabilityof[\vcoeff]{\vobs} \, \probabilityof{\vcoeff},
\end{align}
where the prior \(\probabilityof{\vcoeff}\) is the Gaussian distribution derived above.
We can now estimate the posterior distribution using the \PDMP\ methods described in \cref{sec:pdmp}, i.e. \(\target(\vcoeff) = \probabilityof[\vobs]{\vcoeff}\).
The posterior for the two-dimensional case is visualized in \cref{fig:posterior}.

\subsection{Evaluation Metrics}\label{subsec:evaluation-metrics}
First, we compute a reference solution for each setting \(d=2,\dots,10\) using a long \RWM\ run with \num{e7} samples.
Let \(\mu_i^{\star}\) and \(\sigma_{i}^{2,\star}\) denote the coordinate-wise posterior mean and variance of this reference posterior.
For any method \(\method\), let \(\hat{\mu}^{(\method)}_i\) and \(\hat{\sigma}^{2,(\method)}_i\) denote the corresponding estimates.
We then evaluate the following metrics:\\

\noindent
\textbf{Posterior Mean Error:}
\begin{equation}
    \mathrm{RMSE}_{\mathrm{mean}}^{(\method)}
    = \sqrt{\frac{1}{d}\sum_{i=1}^{d}\bigl(\hat{\mu}^{(\method)}_i - \mu_i^{\star}\bigr)^2}
\end{equation}
\medskip

\noindent
\textbf{Posterior Variance Error:}
\begin{equation}
    \mathrm{RMSE}_{\mathrm{var}}^{(\method)}
    = \sqrt{\frac{1}{d}\sum_{i=1}^{d}\bigl(\hat{\sigma}^{2,(\method)}_i - \sigma_{i}^{2,\star}\bigr)^2}
\end{equation}
\medskip

\noindent
\textbf{Wasserstein Distance:} We compute the 2-Wasserstein distance (WD) between the reference solution \(\hat{p}^*\) and an approximation \(\hat{p}^{(\method)}\) resulting from method \(\method\).
\begin{equation}
    W_2^2(\hat{p}^*,\hat{p}^{(\method)}) = \min_{\pi\in\Pi(\hat{p}^*,\hat{p}^{(\method)})} \iint  \doublepipes{\bapos - \bapos'}^2 \dif{\pi \parens*{\bapos, \bapos'}} \\
    \label{eq:empirical_w2}
\end{equation}
\noindent
where \(\Pi(\hat{p}^*,\hat{p}^{(\method)})\) denotes the set of all couplings of the two distributions.

\medskip

\noindent
\textbf{Effective Sample Size (ESS) per Model Evaluation:} We compute the ESS per model evaluation to assess the sampling efficiency of the \PDMP\ methods.
The ESS is calculated using the formula
\begin{align}
    \text{ESS} = \frac{N}{1 + 2\sum_{k=1}^{\infty} \rho_k},
\end{align}
where \(N\) is the total number of samples and \(\rho_k\) is the autocorrelation at lag \(k\).
We normalize the ESS by dividing it by the number of model evaluations $N_{\mathrm{eval}}$ to account for computational cost.

\bmhead{Remarks} We compute all metrics in the affine space to obtain standardized results.
The posterior mean and variance of the \PDMPs\ can be computed analytically, as shown in~\cite{Bierkens2019b}.
We display these quantities as functions of \PDE\ model evaluations $N_{\mathrm{eval}}$, as we assume these evaluations to be the computational bottleneck of our problem.
This way, we can compare convergence of the different methods on equal footing.
We therefore need to keep track of the number of model evaluations performed by the \PDMP\ samplers.
Markov chains are typically evaluated in terms of number of samples drawn, which is directly proportional to the number of model evaluations for the \RWM\ sampler.

In~\cite{Bierkens2019b}, a method to compute the ESS directly from the \PDMP\ skeleton without a need to discretize is also provided.
However, we observe in the shorter runs considered here that the ESS computed from the discretized process produces more stable and accurate results.
We therefore approximate the \PDMP\ until time \(T\) and discretize it into \(N\) samples equally spaced in time, i.e. \(\bapos_i = \bapos(i \cdot T / N)\) for \(i = 1, \ldots, N\).

In a similar manner, we discretize the PDMPs for the empirical WD, as no analytical solution for this metric is available.
The exact empirical WD becomes prohibitively expensive to compute for large sample sizes and dimensions.
We therefore resort to the debiased Sinkhorn divergence, as described in~\cite{Feydy2018}, as an efficient approximation.
We set regularization parameter \(\varepsilon = 0.02\) for numerical stability while keeping the bias low.

The WD is also reported as a function of model evaluations, similar to the posterior mean and variance.
The empirical WD is a biased and noisy estimator of the true WD between the \PDMP\ and the reference posterior, especially for small sample sizes.
We do not correct for the bias in our experiments, but ensure comparability by using the same number of samples for all methods.
This means that we discretize the PDMPs with \(N\) points when computing the WD distance after \(N\) model evaluations.
We reduce the variance by taking multiple random draws of \(N\) samples from the reference posterior.

We repeat each numerical experiment 50 times with 50 different seeds, and then report the average for each of the four metrics.
For each seed, the initial position of the sampling algorithm is drawn from the Laplace approximation of the posterior.
Note that we run the MCMC algorithms in the affine space as well to put all algorithms on an equal footing.
This way, we expect a short burn-in phase regardless of the algorithm.
We obtain the affine transformation only once per setting and use it for all methods and seeds.

We provide a RWM baseline for all experiments, which is not to be confused with the reference solution with \num{e7} samples.
It is well known that the performance of RWM depends heavily on the choice of the proposal distribution.
We start out with a standard Gaussian proposal, reflecting the isotropic nature of the posterior in the affine space.
After each \num{100} iterations, we multiply the current proposal covariance with \num{0.9} if the acceptance rate is below \num{0.2}, and with \num{1.1} if the acceptance rate is above \num{0.25}.
After \num{1000} iterations, we set the proposal covariance matrix to the empirical covariance of the samples collected so far, scaled by the optimal factor \(\nicefrac{2.38^2}{d}\) derived in~\cite{Gelman1997}.
We then keep rescaling the proposal covariance every \num{100} iterations as before, and leave it unchanged after \num{2000} iterations to satisfy vanishing adaptation~\cite{Andrieu2008}.
It is also this version of the \RWM\ algorithm that we use to compute the reference solutions, albeit with a different seed and a much longer run where we discard the first \num{5000} samples as burnin. \subsection{Convergence of the Zig-Zag Sampler}\label{subsec:when_converge}
First, we would like to elucidate when the \ZZS\ converges to the true posterior distribution.
This is particularly important as we move beyond established means to simulate the process that are well studied and known to converge.
There is a number of potential violations of the assumptions that guarantee convergence:
\begin{enumerate}
    \item The surrogate model does not produce a strict upper bound process for the thinning.
    \item We constantly change the offset \(\deltarate\) to restore the upper bound property (although to a lesser extent at later times).
    \item We constantly reduce the offset to recover from a potentially large gap between the surrogate and the true model.
    \item We revert and redo events when the surrogate is found to be an invalid upper bound and the offset \(\deltarate\) is increased.
\end{enumerate}
We compare the following three scenarios to investigate the influence of these potential violations on performance and accuracy of the sampler:
\begin{itemize}
    \item \textbf{Constant potential:} We use a constant surrogate model and increase the offset \(\deltarate\) when needed.
    Additionally, the offset shrinks exponentially with time \(t\).
    This tests points 2, 3, and 4.
    \item \textbf{Constant potential with no shrinkage:} We use a constant surrogate model and increase the offset \(\deltarate\) when needed, but do not allow it to shrink.
    This tests points 2 and 4.
    \item \textbf{Random gradient:} We use a surrogate model that produces a random value for each iteration of the \ZZS.
    We sample a value from \( \uniform{-0.5}{0.5}\) for each component of the potential gradient.
    We increase the offset \(\deltarate\) when needed, and allow it to shrink.
    This tests the limits of admitting a non-static upper bound process, \ie tests points 1, 2.
\end{itemize}
The two constant surrogate models represent the most naive approach to surrogate modelling, where we simply use a constant value as an upper bound for the rate function.
In theory, such a model would only be applicable to bounded domains, or a small number of distributions on unbounded domains, such as the Cauchy distribution.
A Gaussian distribution, \eg produces an unbounded rate function.
In practice, however, all realizations of the \ZZS\ are finite in time, and thus the rate function is bounded along any simulated  trajectory.
We would like to find this maximum value of the rate function with this constant model.

The random surrogate model represents a highly inaccurate model that does not provide much useful information about the rate function.
Further, it is an extreme case of an ever-changing upper bound, which helps to isolate the influence of tampering with the upper bound process over time.
\begin{figure}
    \centering
    \subcaptionbox{\(d=2\) \label{fig:when_converge_2d_rmse}}[0.33\textwidth]{\includegraphics[width=\linewidth]{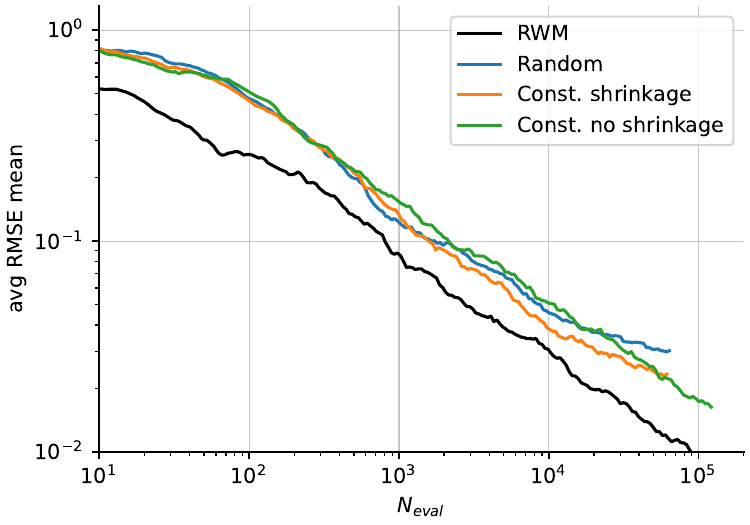}}\subcaptionbox{\(d=5\) \label{fig:when_converge_5d_rmse}}[0.33\textwidth]{\includegraphics[width=\linewidth]{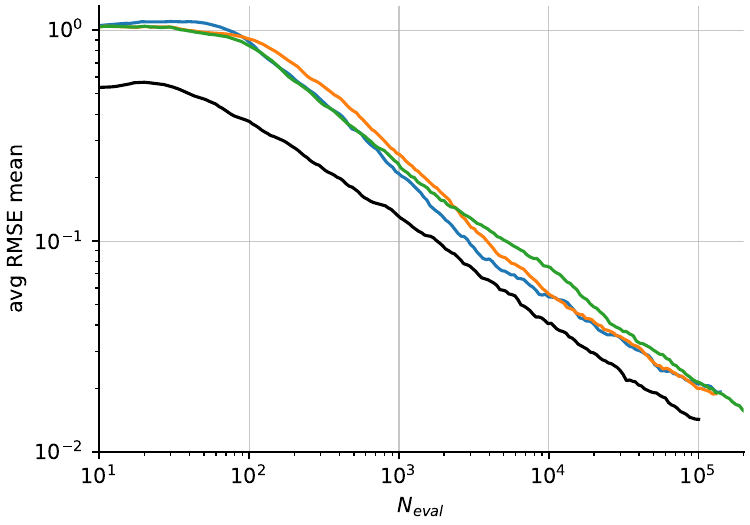}}\subcaptionbox{\(d=10\) \label{fig:when_converge_10d_rmse}}[0.33\textwidth]{\includegraphics[width=\linewidth]{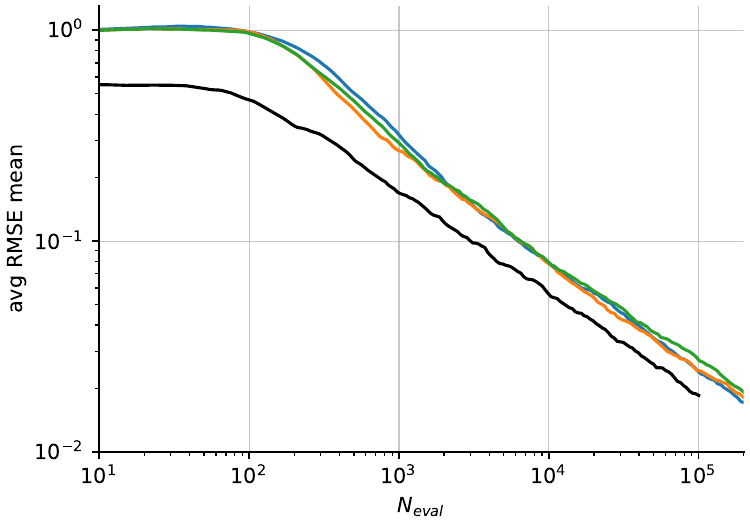}}\\
    \subcaptionbox{\(d=2\) \label{fig:when_converge_2d_var}}[0.33\textwidth]{\includegraphics[width=\linewidth]{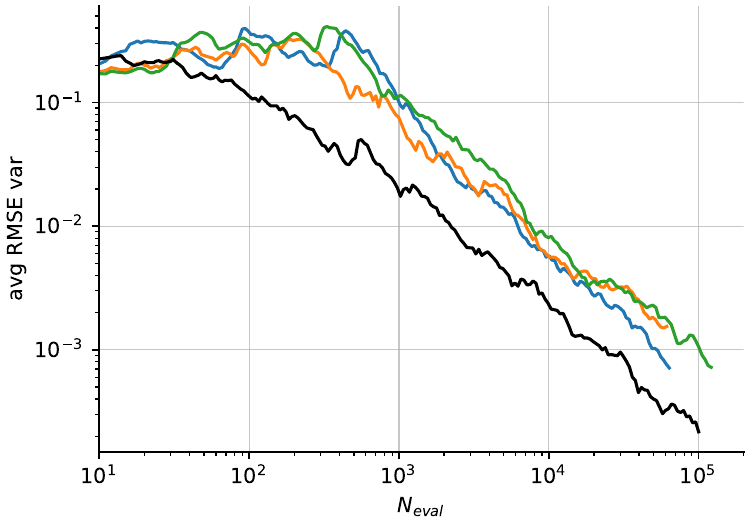}}\subcaptionbox{\(d=5\) \label{fig:when_converge_5d_var}}[0.33\textwidth]{\includegraphics[width=\linewidth]{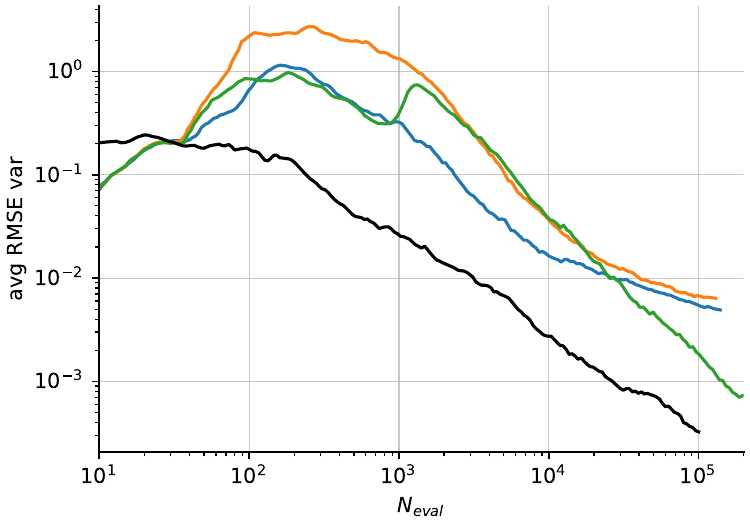}}\subcaptionbox{\(d=10\) \label{fig:when_converge_10d_var}}[0.33\textwidth]{\includegraphics[width=\linewidth]{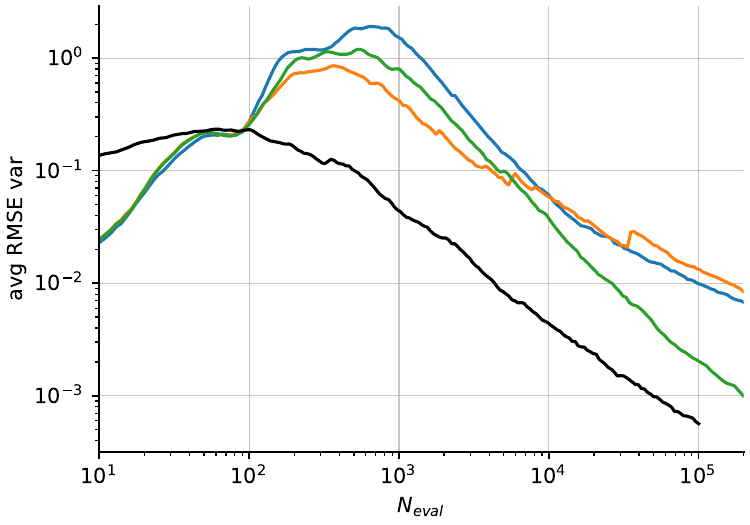}}\caption{
        \RMSE\ of the mean (first row) and variance (second row) as functions of the number of model evaluations.
        All models converge, but perform worse than the baseline \RWM\ sampler.
    }
    \label{fig:when_converge}
\end{figure}

\begin{figure}
    \centering
    \includegraphics[width=.4\linewidth]{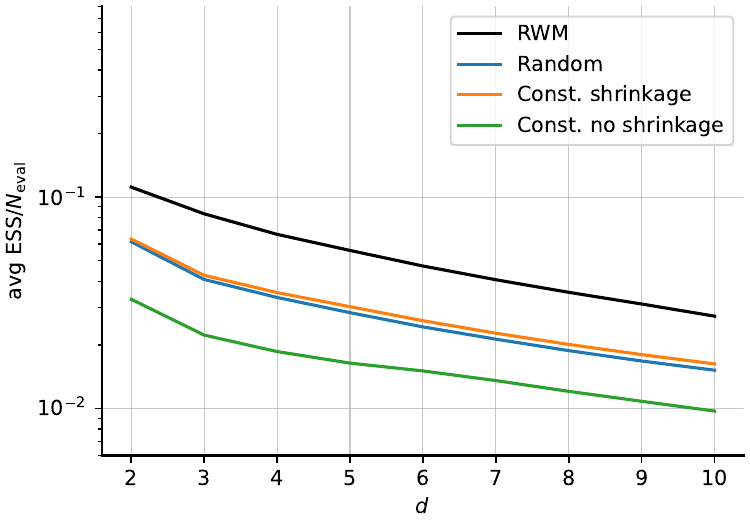}
    \caption{Average \ESS\ per model evaluation as a function of the dimensionality.}
    \label{fig:when_converge_2d_average_ess}
\end{figure}
In \cref{fig:when_converge}, we compare the three scenarios in terms of the average \RMSE\ of the posterior mean (top) and variance (bottom) estimates in \(d=2\), \(d=5\), and \(d=10\) dimensions.
We also report the average \ESS\ per model evaluation in \cref{fig:when_converge_2d_average_ess}.
The first thing to point out is that all three scenarios converge to the true posterior distribution, as indicated by the \RMSE\ of the posterior mean estimate in~\cref{fig:when_converge_2d_rmse,fig:when_converge_5d_rmse,fig:when_converge_10d_rmse}.
We can also observe convergence for the posterior variance in~\cref{fig:when_converge_2d_var,fig:when_converge_5d_var,fig:when_converge_10d_var}, albeit with slight upward curvatures at later stages for the models with shrinkage.
This is an important result, as it indicates that the \ZZS\ is robust to the potential violations of the assumptions that guarantee convergence, and in line with~\cite{Corbella2022}, where a new upper bound is found for each event.

The second thing to note is the constant model with shrinkage performing best in terms of \ESS\ per model evaluation, with a slight edge over the random model with shrinkage, as shown in \cref{fig:when_converge_2d_average_ess}.
They both comfortably outperform the constant model without shrinkage in this metric.
The frontrunner in the constant model is not surprising, as it should produce the tightest upper bound out of the three.
The runner-up being the random model is less intuitive.
This finding suggests two things:
i) While the randomness in the model might lead to an increase of the offset, it could by chance produce a tighter upper bound in regions of low process rate.
ii) Shrinkage of the offset matters with such a naive surrogate model.
More generally, we find that the influence of the shrinkage parameter \( \decay \) is less pronounced with better surrogate model accuracy, as discussed in detail in~\ref{sec:influence_shrinkage}.
The key takeaway from the analysis is that as long as the surrogate model is sufficiently accurate, the \ZZS\ is robust to the choice of the shrinkage parameter.

Finally, it is obvious that the three models are not sufficiently accurate, as the \ZZS\ is outperformed by the \RWM\ sampler in terms of \RMSE\ and \ESS\ per model evaluation in all dimensions.
This clearly illustrates the need for better surrogate models.

\subsection{Surrogate Model Comparison}\label{subsec:surrogate_model_comparison}
We now compare the performance of different surrogate models in terms of convergence rates and sampling efficiency.
We consider the constant surrogate model, the Laplace approximation, and two Gaussian process surrogate models:
one that observes only the function values, and one that additionally observes the gradients (Grad \GPabbr).
Both \GPabbr\ models are trained with \num{25} points per latent space dimension.
We set the shrinkage parameters to \(\decay = \num{2e-2}\) for all simulations.

\begin{figure}
    \centering
    \subcaptionbox{\(d=2\) \label{fig:surrogates_2d_mean}}[0.33\textwidth]{\includegraphics[width=\linewidth]{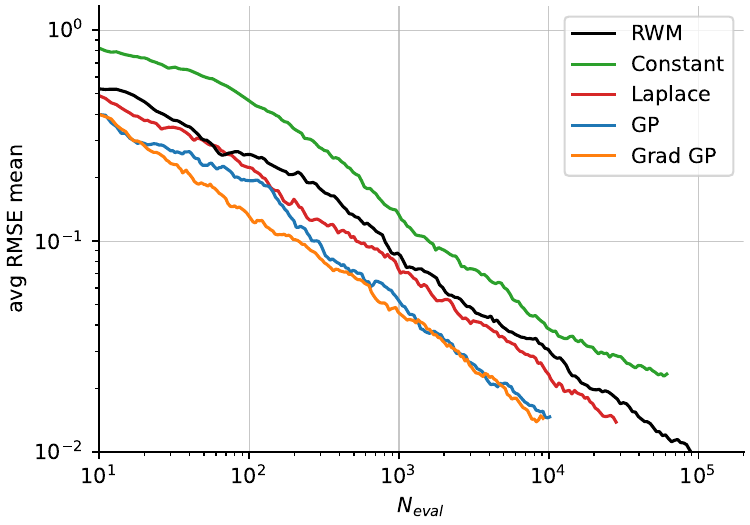}}\subcaptionbox{\(d=5\) \label{fig:surrogates_5d_mean}}[0.33\textwidth]{\includegraphics[width=\linewidth]{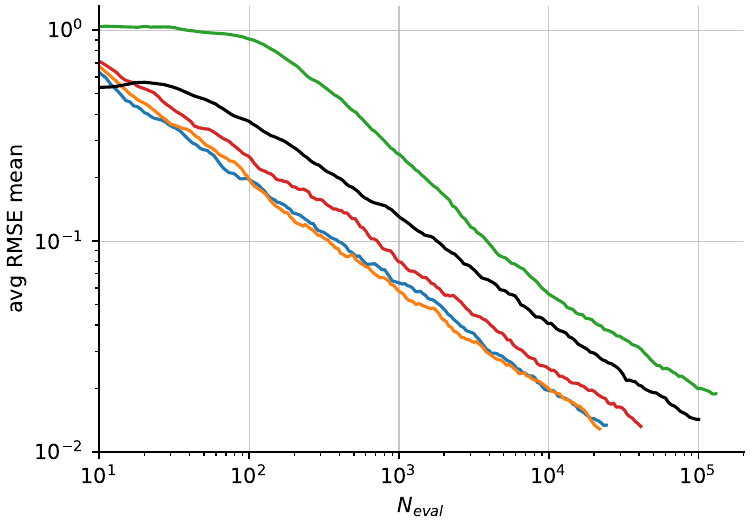}}\subcaptionbox{\(d=10\) \label{fig:surrogates_10d_mean}}[0.33\textwidth]{\includegraphics[width=\linewidth]{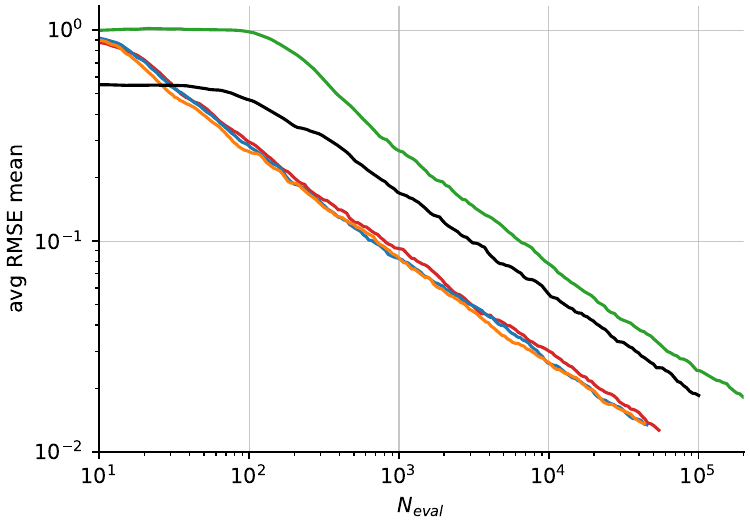}}\\
    \subcaptionbox{\(d=2\) \label{fig:surrogates_2d_var}}[0.33\textwidth]{\includegraphics[width=\linewidth]{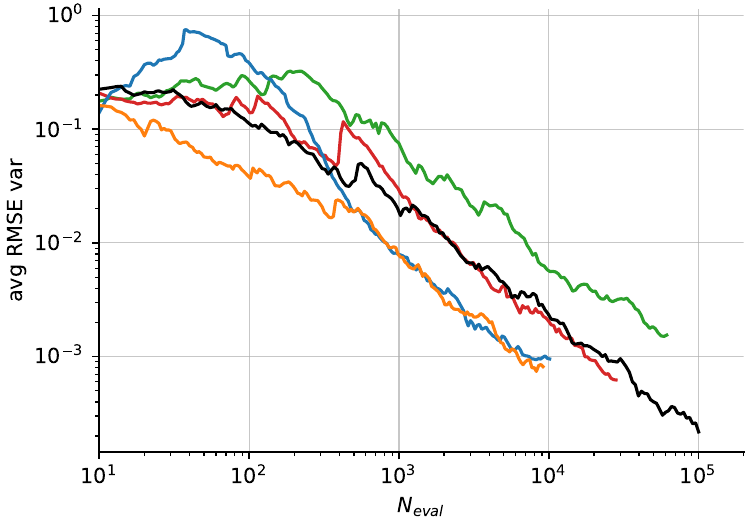}}\subcaptionbox{\(d=5\) \label{fig:surrogates_5d_var}}[0.33\textwidth]{\includegraphics[width=\linewidth]{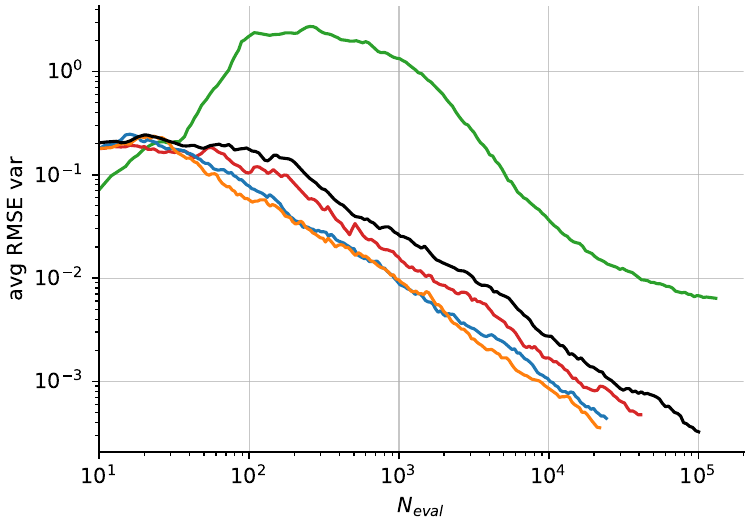}}\subcaptionbox{\(d=10\) \label{fig:surrogates_10d_var}}[0.33\textwidth]{\includegraphics[width=\linewidth]{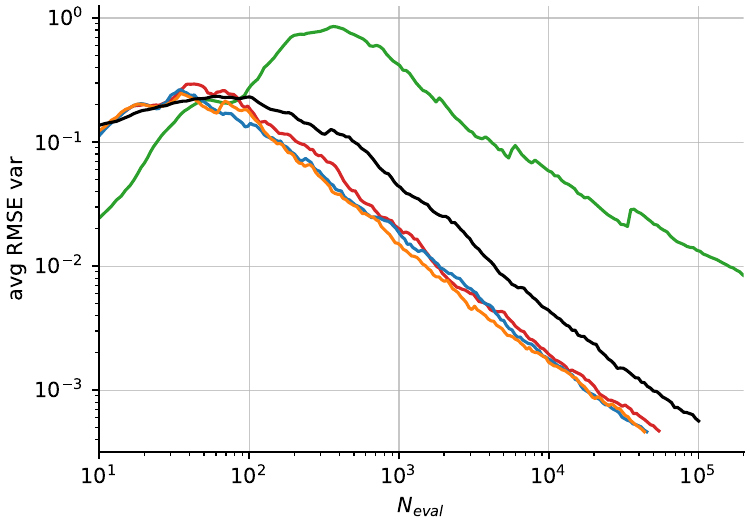}}\caption{
        Comparison of surrogate models in terms of \RMSE\ of the mean (first row) and \RMSE\ of the variance (second row).
        The more flexible Laplace, \GPabbr, and Grad \GPabbr\ models are accurate enough to outperform the \RWM\ sampler.
    }
    \label{fig:surrogates}
\end{figure}

\Cref{fig:surrogates} shows the average \RMSE\ of the posterior mean (top row) and variance (bottom row) estimates for \(d=2\), \(d=5\), and \(d=10\) dimensions.
With all surrogates except the constant model, the \ZZS\ outperforms the \RWM\ sampler across all metrics and dimensionalities.
This even holds for the Laplace model, which is remarkable, as it is a rather crude approximation of the rate function.
The rest of the models complete a consistent hierarchy, with the Grad \GPabbr\ performing best, followed by the \GPabbr, the Laplace, and finally the constant model.
However, the Grad \GPabbr\ has the upper hand only by a small margin, suggesting that the additional information from the gradients is not crucial in these settings.
In fact, the margin between the non-constant models shrinks with increasing dimensionality, indicating that the target distribution becomes more isotropic with finer discretizations of the underlying field.

\begin{figure}
    \centering
    \includegraphics[width=0.4\linewidth]{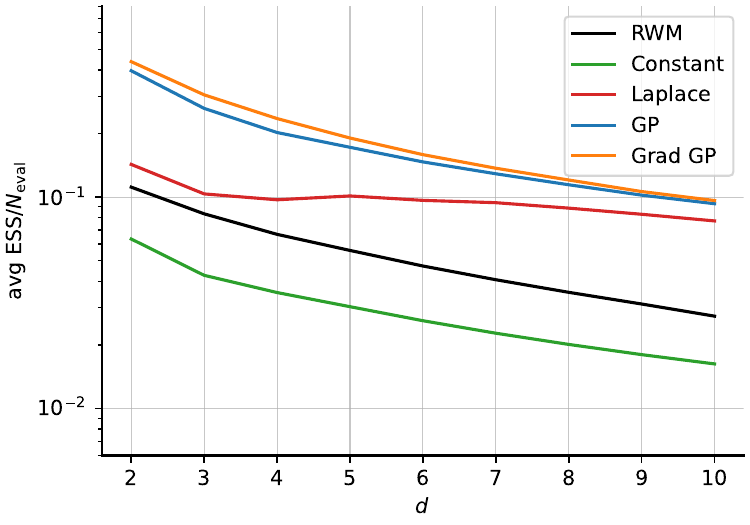}
    \caption{
        Average \ESS\ per model evaluation as a function of the dimensionality for all surrogate models.
        All but the constant surrogate outperform the \RWM\ sampler.
    }
    \label{fig:surrogates_all_ess}
\end{figure}

\Cref{fig:surrogates_all_ess} shows the average \ESS\ per model evaluation as a function of the dimensionality for all surrogate models.
The ranking of the models is consistent with the \RMSE\ results, with the Grad \GPabbr\ leading the field, followed by the \GPabbr, the Laplace, the \RWM\ sampler, and finally the constant model.

\subsection{Adaptive Refinement of the \GPabbr\ Surrogate Model}\label{subsec:active_learning}
\begin{figure}[H]
    \centering
    \subcaptionbox{\(d=2\) \label{fig:adaptive_2d_mean}}[0.33\textwidth]{\includegraphics[width=\linewidth]{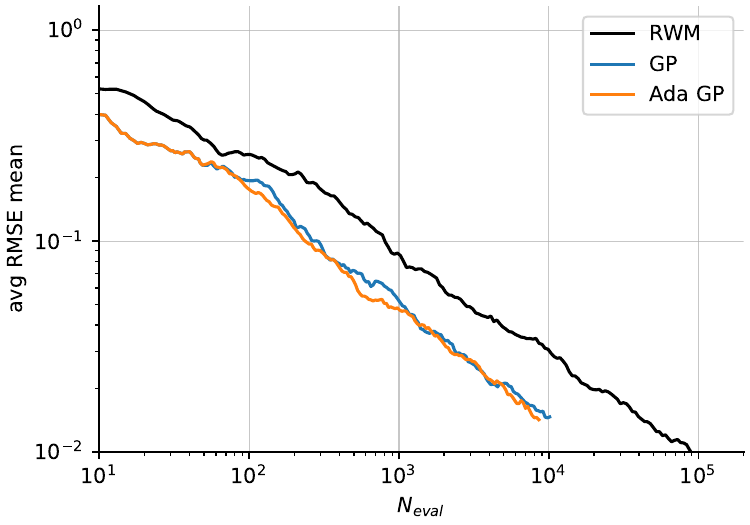}}\subcaptionbox{\(d=5\) \label{fig:adaptive_5d_mean}}[0.33\textwidth]{\includegraphics[width=\linewidth]{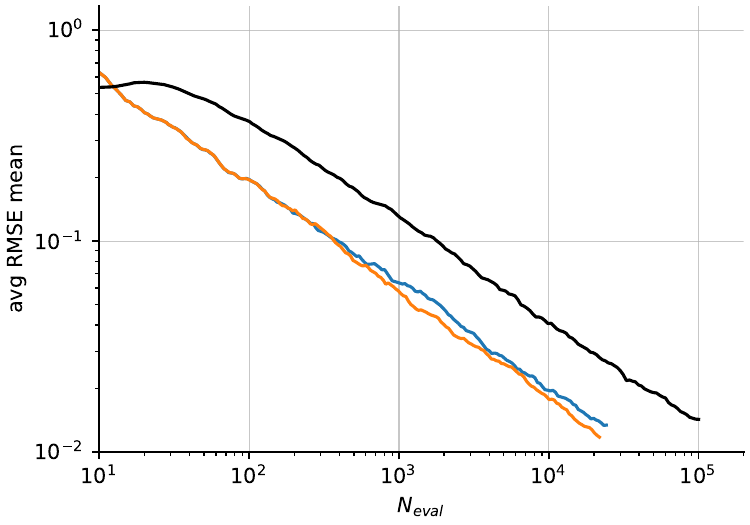}}\subcaptionbox{\(d=10\) \label{fig:adaptive_10d_mean}}[0.33\textwidth]{\includegraphics[width=\linewidth]{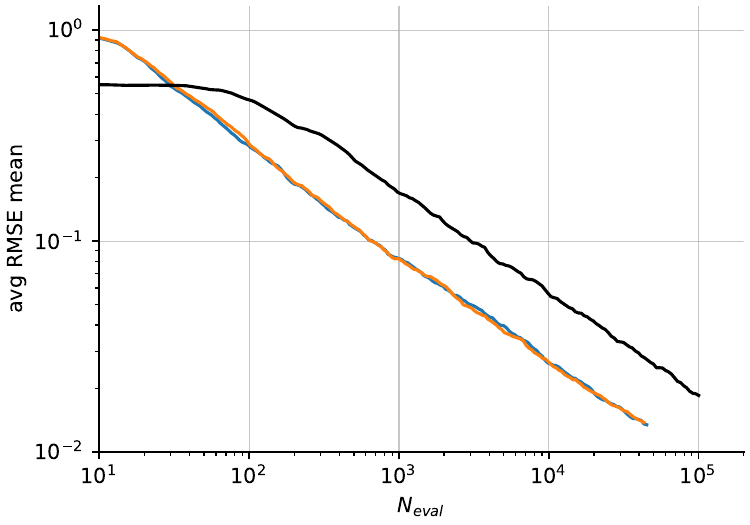}}\\
    \subcaptionbox{\(d=2\) \label{fig:adaptive_2d_var}}[0.33\textwidth]{\includegraphics[width=\linewidth]{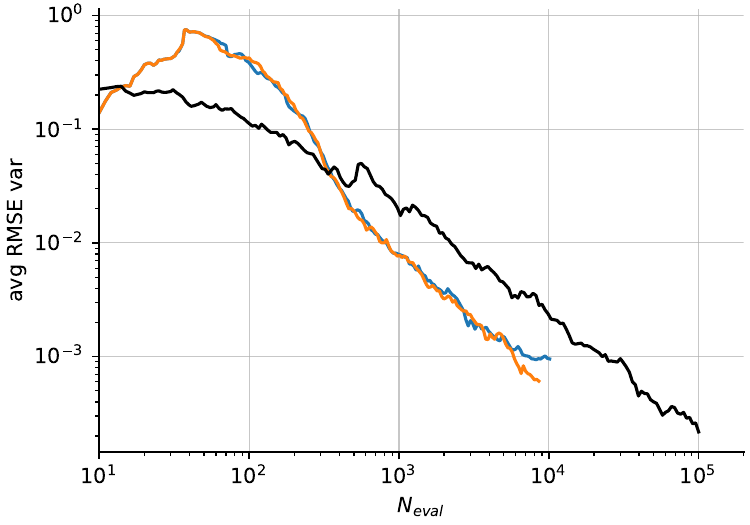}}\subcaptionbox{\(d=5\) \label{fig:adaptive_5d_var}}[0.33\textwidth]{\includegraphics[width=\linewidth]{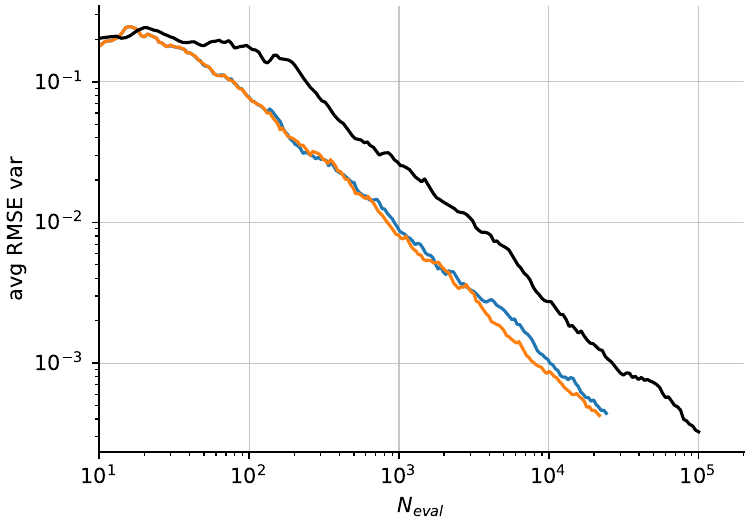}}\subcaptionbox{\(d=10\) \label{fig:adaptive_10d_var}}[0.33\textwidth]{\includegraphics[width=\linewidth]{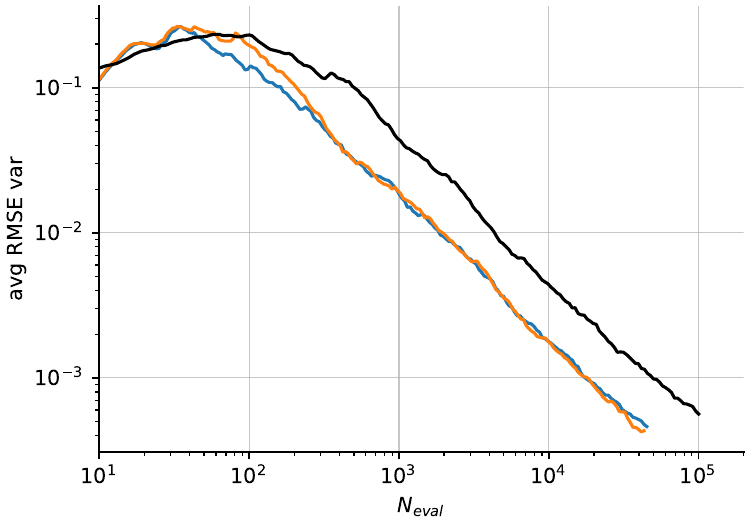}}\caption{
        Comparison of pretrained \GPabbr\ and Adaptive \GPabbr\ in terms of \RMSE\ of the mean (first row) and \RMSE\ of the variance (second row).
        The addition of more training points during the simulation does not consistently improve performance.
    }
    \label{fig:adaptive}
\end{figure}

The last extension to the surrogate model is to adaptively refine the surrogate model during the \PDMP\ simulation.
Since we need to evaluate the model at each proposed event time, we can use the additional information from these points to update the surrogate model.
We compare the pretrained \GPabbr\ surrogate model with an adaptive \GPabbr\ version (Ada \GPabbr) in \cref{fig:adaptive}.

\begin{figure}
    \centering
    \includegraphics[width=0.4\linewidth]{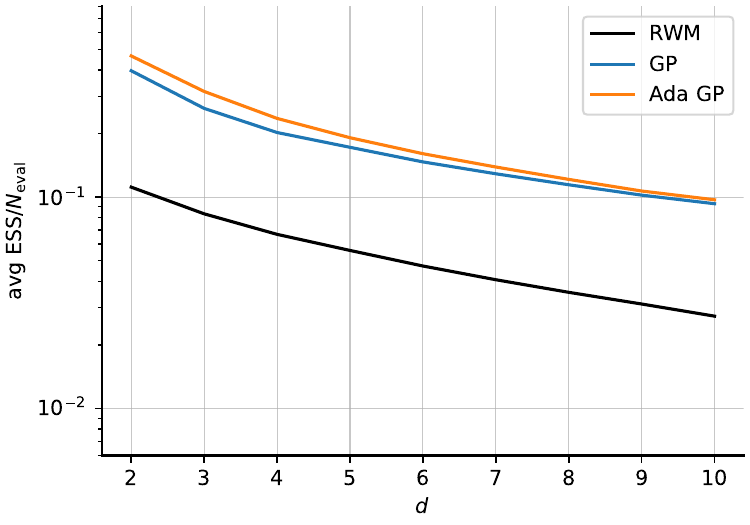}
    \caption{Average \ESS\ per model evaluation as a function of the dimensionality for the \GPabbr\ and Ada \GPabbr\ models.}
    \label{fig:adaptive_2d_average_ess}
\end{figure}

The Ada \GPabbr\ and \GPabbr\ curves fully overlap at the start of the simulation, as the Ada \GPabbr\ model is initialized with the same training data as the \GPabbr\ model.
The adaptive refinement becomes apparent in later stages of the simulation, where the \RMSE\ curves of the two models become distinct.
Counterintuitively, the Ada \GPabbr\ does not consistently outperform the \GPabbr\ model.
In fact, the two curves cross multiple times in all three metrics.
The \ESS\ per model evaluation is slightly higher for the Ada \GPabbr\ model than for the \GPabbr\ models, but only by the smallest of margins, as shown in \cref{fig:adaptive_2d_average_ess}.

We hypothesize that the additional training points added during the simulation lie in regions well covered by the initial training set, and thus do not contribute much new information to the model.
While the initial sampling from the Laplace approximation yields high-quality training points for this problem with moderate non-linearity, the situation may differ when strong non-linearities are present.
We therefore assume that the benefits of adaptive refinement are strongly problem-dependent, and that more sophisticated adaptive sampling strategies are required to fully leverage its potential in more challenging scenarios.

\subsection{Zig-Zag Sampler vs Bouncy Particle Sampler vs No-U-Turn Sampler}\label{subsec:zz_vs_bps_vs_nuts}
\begin{figure}
    \centering
    \subcaptionbox{ZZS\label{subfig:zz}}{\includegraphics[width=0.32\linewidth]{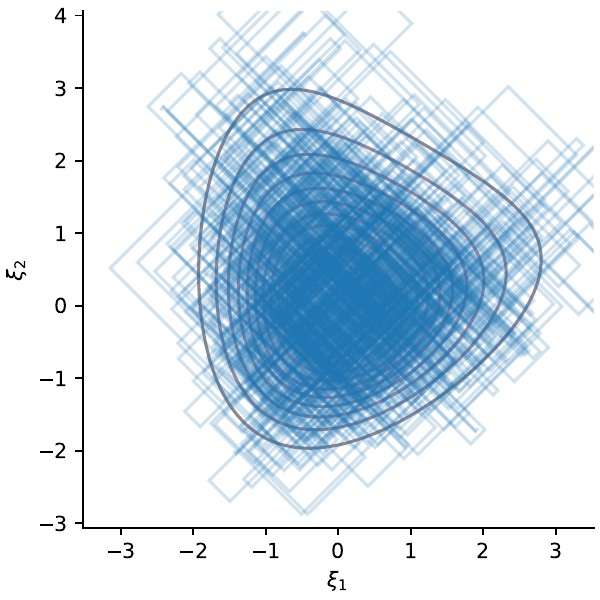}}\subcaptionbox{BPS\label{subfig:bps_0-1}}{\includegraphics[width=0.32\linewidth]{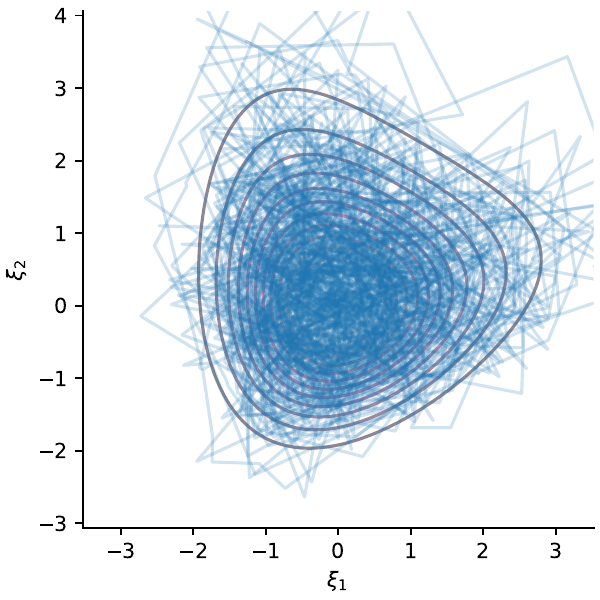}}\subcaptionbox{NUTS\label{subfig:bps_0}}{\includegraphics[width=0.32\linewidth]{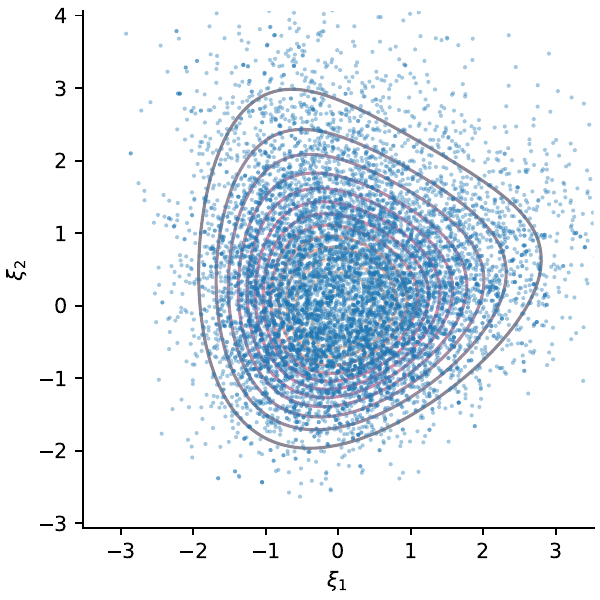}}\caption{Path of the (a) Zig-Zag sampler and (b) Bouncy Particle sampler, and (c) samples from the \NUTS.}\label{fig:zz_bps_path}
\end{figure}

Finally, we compare the performance of the \ZZS\ with the \BPS\ and the \NUTS\@.
This comparison is necessary, as the latter is considered the state-of-the-art \HMC\ algorithm for continuous distributions.
For the \ZZS\ and \BPS, we use the \GPabbr\ surrogate model with 25 points per latent space dimension.
For \NUTS, we use our own implementation of the algorithm with dual-averaging of the step size, as described in~\cite{Hoffman2014}.
We cold-start the \ZZS\ and the \NUTS, \ie we run them without any tweaking of the hyperparameters ahead of the inference.
For \BPS, we use the refreshment rate deemed suitable in \cref{subsec:bps_refreshment_rate}, \ie \(\refreshrate = \num{e-1}\).
We show the paths of the \ZZS, the \BPS, and the \NUTS\ in \cref{fig:zz_bps_path}.

\begin{figure}
    \centering
    \subcaptionbox{\(d=2\) \label{fig:zz_bps_nuts_2d_mean}}[0.33\textwidth]{\includegraphics[width=\linewidth]{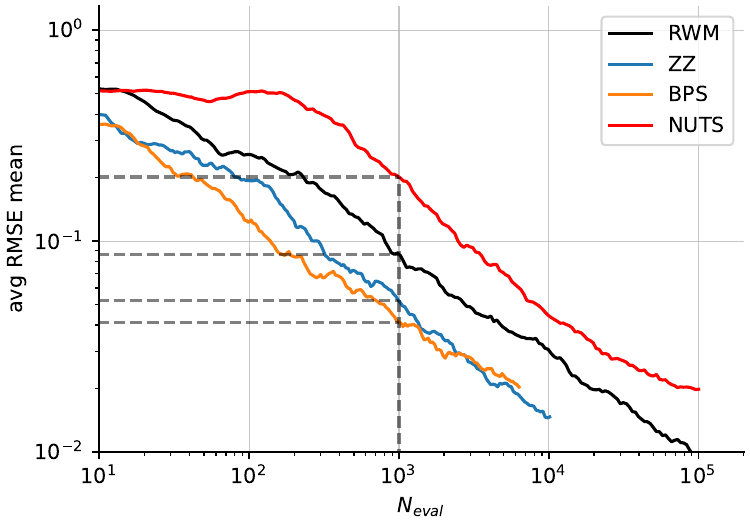}}\subcaptionbox{\(d=5\) \label{fig:zz_bps_nuts_5d_mean}}[0.33\textwidth]{\includegraphics[width=\linewidth]{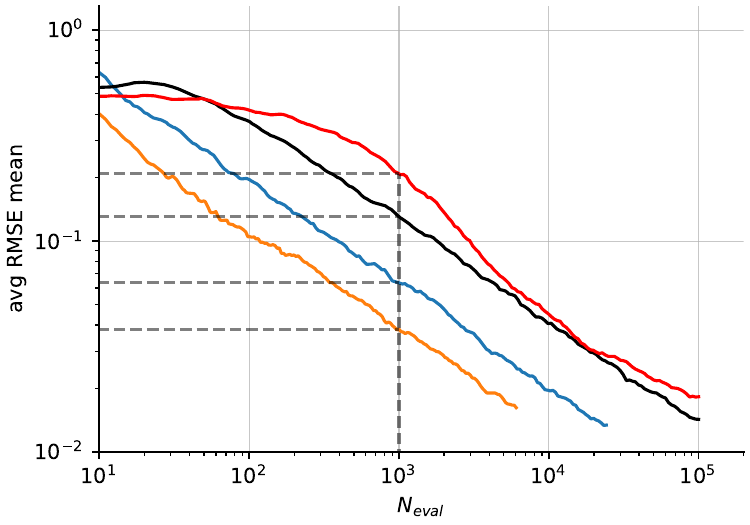}}\subcaptionbox{\(d=10\) \label{fig:zz_bps_nuts_10d_mean}}[0.33\textwidth]{\includegraphics[width=\linewidth]{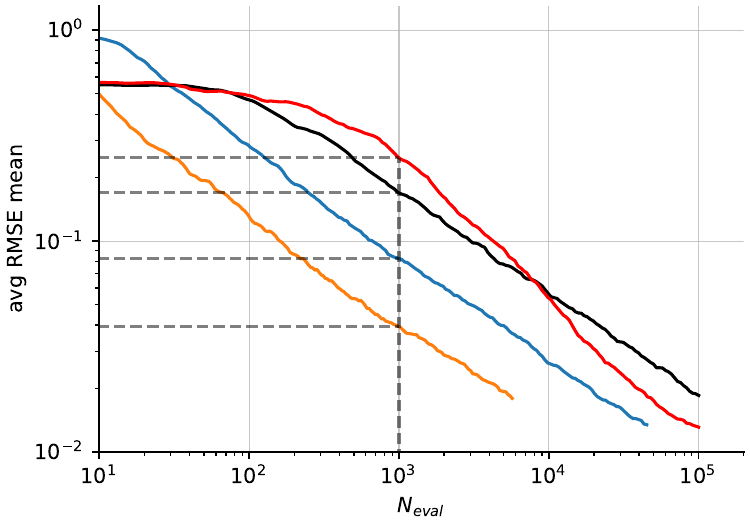}}\\
    \subcaptionbox{\(d=2\) \label{fig:zz_bps_nuts_2d_var}}[0.33\textwidth]{\includegraphics[width=\linewidth]{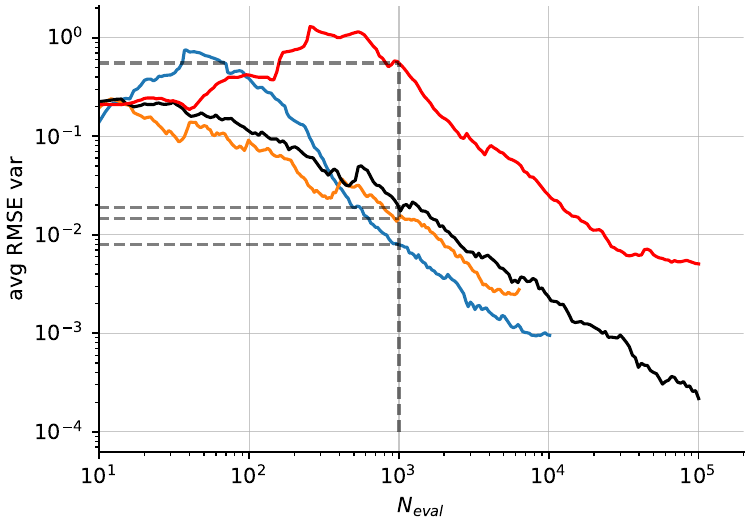}}\subcaptionbox{\(d=5\) \label{fig:zz_bps_nuts_5d_var}}[0.33\textwidth]{\includegraphics[width=\linewidth]{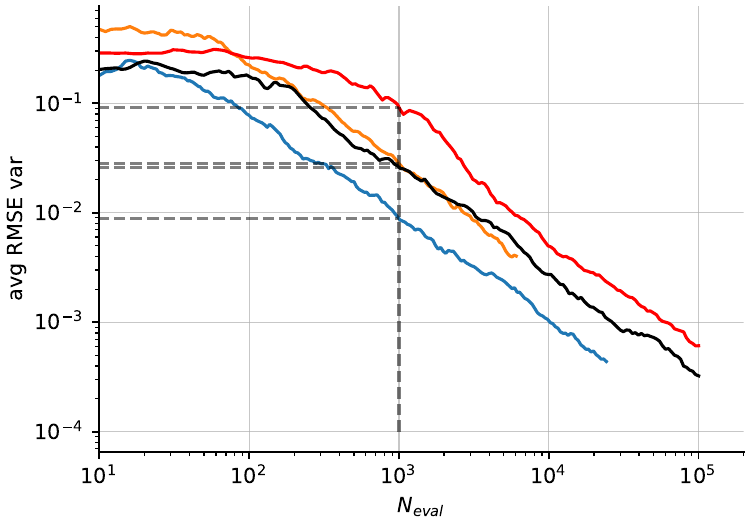}}\subcaptionbox{\(d=10\) \label{fig:zz_bps_nuts_10d_var}}[0.33\textwidth]{\includegraphics[width=\linewidth]{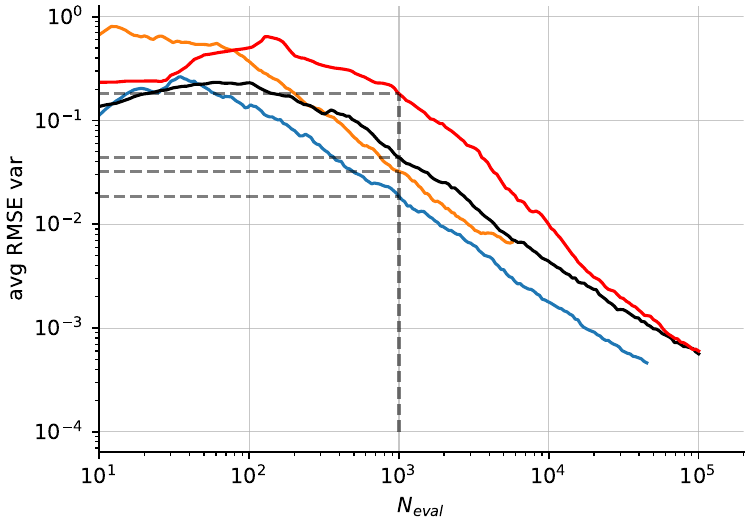}}\\
    \subcaptionbox{\(d=2\) \label{fig:zz_bps_nuts_2d_wd}}[0.33\textwidth]{\includegraphics[width=\linewidth]{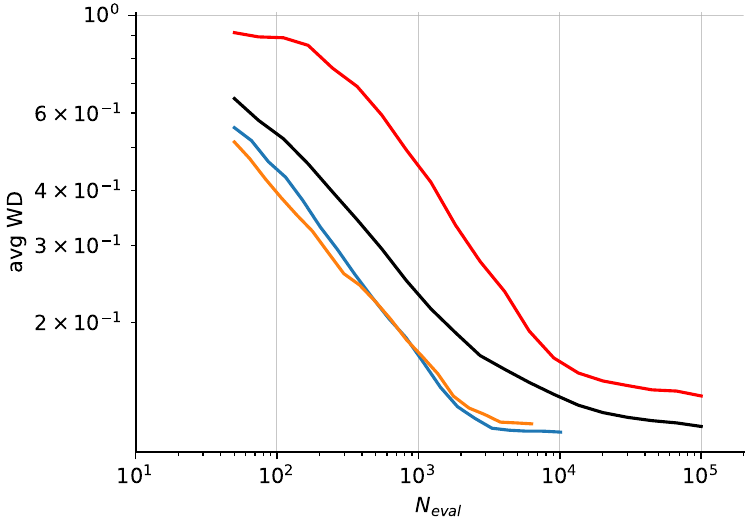}}\subcaptionbox{\(d=5\) \label{fig:zz_bps_nuts_5d_wd}}[0.33\textwidth]{\includegraphics[width=\linewidth]{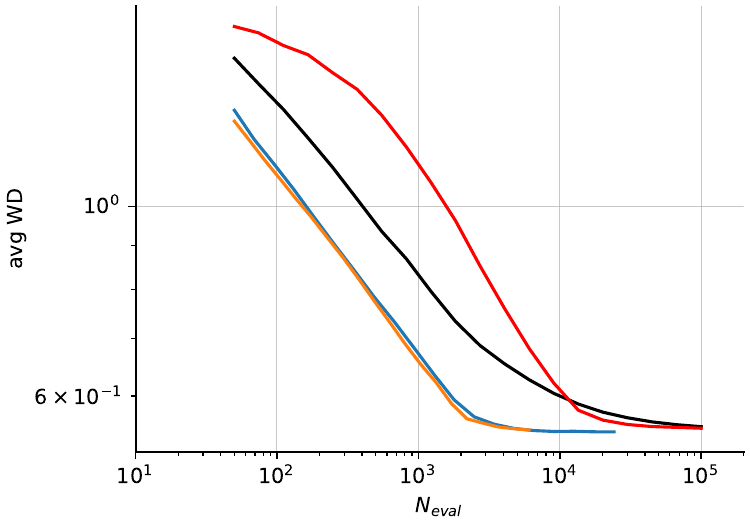}}\subcaptionbox{\(d=10\) \label{fig:zz_bps_nuts_10d_wd}}[0.33\textwidth]{\includegraphics[width=\linewidth]{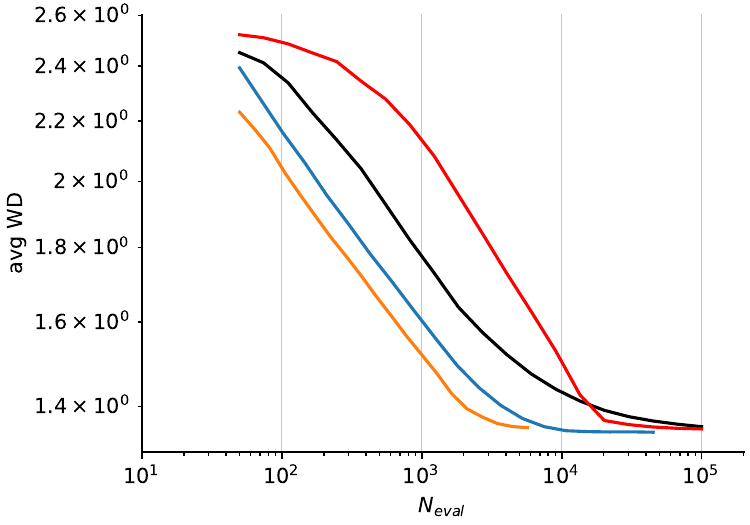}}\caption{
        Comparison of \ZZS, \BPS, and \NUTS\ in terms of \RMSE\ of the mean (top row), \RMSE\ of the variance (middle row), and \WD\ (bottom row).
        Both \PDMPs\ outperform the \NUTS\ in all dimensions regarding the posterior mean and variance estimates and the WD.
    }
    \label{fig:zz_bps_nuts}
\end{figure}

The \BPS\ clearly outperforms the \ZZS\ in terms of the posterior mean estimate, as shown in \cref{fig:zz_bps_nuts_2d_mean,fig:zz_bps_nuts_5d_mean,fig:zz_bps_nuts_10d_mean}.
Both of the \PDMP\ algorithms outperform the \NUTS\ in all dimensions regarding this metric:
Assuming a budget of \num{1000} model evaluations, both the \ZZS\ and the \BPS\ achieve a \RMSE\ of about \num{0.04}, whereas the \NUTS\ only reaches about \num{0.085} for \(d=2\)\@.
In \(d=10\) dimensions, the \BPS\ achieves a \RMSE\ of about \num{0.04}, whereas the \NUTS\ is at \num{0.2}, with the \ZZS\ in between at about \num{0.08}.
The \NUTS\ shows the worst performance in all dimensions, producing a \RMSE\ of about \num{0.2}.

A slightly different picture emerges when looking at the posterior variance estimates, as shown in \cref{fig:zz_bps_nuts_2d_var,fig:zz_bps_nuts_5d_var,fig:zz_bps_nuts_10d_var}:
Here, the \ZZS\ outperforms both the \BPS\ and \NUTS\ in all dimensions.
The \BPS\ still outperforms the \NUTS, but produces an equally accurate variance estimate as the \RWM\ algorithm.

As no algorithm dominates across both metrics, we also consider the \WD\ between the estimates from the different samplers and the reference solution.
Inspecting the \WD\ in \cref{fig:zz_bps_nuts_2d_wd,fig:zz_bps_nuts_5d_wd,fig:zz_bps_nuts_10d_wd}, we see that both \PDMPs\ perform similarly well in \(d=2\) and \(d=5\) dimensions, with a slight edge for the \BPS\ in the \(d=10\) case.
The \NUTS\ again performs worst in all dimensions.
However, it should be noted that the \NUTS\ is structurally disadvantaged in this comparison:
for a given budget of \(N_{\mathrm{eval}}\) model evaluations it returns fewer posterior samples than the \PDMP\ samplers (due to multiple leapfrog steps per accepted trajectory).
The \WD\ is thus computed from a smaller number of samples and exhibits a higher finite-sample bias.

\begin{figure}
    \centering
    \includegraphics[width=0.4\linewidth]{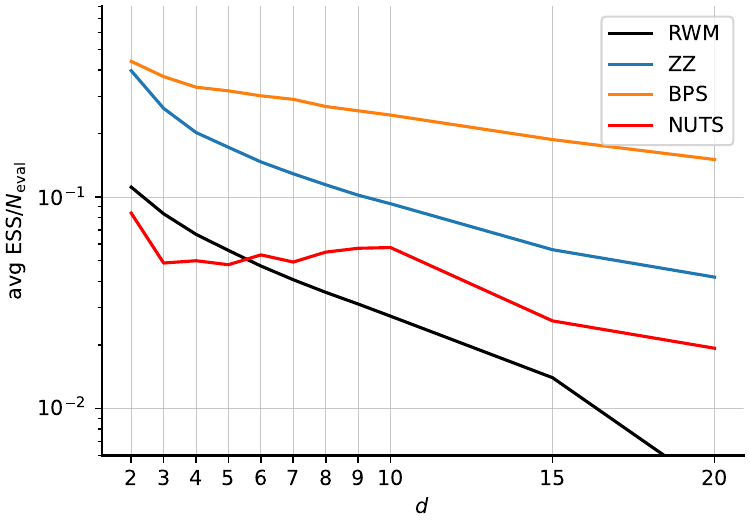}
    \caption{
        Average \ESS\ per model evaluation as a function of the dimensionality for \ZZS, \BPS, \NUTS, and \RWM\@.
        The \BPS\ is the most efficient sampler, followed by the \ZZS, \NUTS, and \RWM.
    }
    \label{fig:zz_bps_nuts_2d_average_ess}
\end{figure}

A similar picture emerges when looking at the \ESS\ per model evaluation, as shown in \cref{fig:zz_bps_nuts_2d_average_ess}:
The \BPS\ is leading the field, followed by the \ZZS, and finally the \NUTS\@.
The \NUTS\ only beats the \RWM\ sampler for \(d>6\), thanks to its favorable scaling with dimensionality.
The slope of the \ZZS\ is similar to the one of the \RWM\ algorithm.
The \BPS, however, displays a slop closer to the one of the \NUTS, albeit with a higher offset, suggesting better scaling with dimensionality than the \ZZS\@.

At a first glance, the weak performance of the \NUTS\ sampler in this comparison is surprising, given its reputation as a state-of-the-art \MCMC\ algorithm.
There are several possible explanations for this outcome:
i) The affine transformation leads to a more isotropic distribution, which improves the performance of the \ZZS\ and the \RWM\ algorithm relative to the \NUTS\@.
ii) The distribution only shows moderate non-linearities.
iii) The cold-start of the \NUTS\ sampler leads to suboptimal hyperparameters, which hamper its performance.
In many practical applications, users would invest \(>\num{e3}\) iterations in tuning the hyperparameters of \NUTS\ ahead of the actual inference.
When expensive PDE model evaluations are involved, such an upfront investment is typically not feasible.
Simply put, the \NUTS\ does not excel in the settings considered here, even though it does in many others.

  \section{Conclusions}\label{sec:concusions}

\noindent
In this work, we introduced and analyzed a surrogate–assisted framework for simulating piecewise-deterministic Markov process (\PDMP) samplers in non-linear Bayesian inverse problems governed by partial differential equations (\PDE s).
Focusing on the Zig-Zag sampler (\ZZS) and the Bouncy Particle sampler (\BPS), we designed a global thinning strategy that uses a surrogate for the potential together with a simple, yet effective, correction mechanism for the resulting event rate.
This construction makes the method agnostic to the particular surrogate family: whenever the surrogate underestimates the true rate locally, the offset is increased to restore the upper–bound property required for valid thinning.

Our numerical experiments---the inference of a coefficient field in linear elasticity---demonstrated that this approach yields \PDMP\ samplers that were both robust and efficient.
Even very crude surrogates, such as a constant model or a pure Laplace approximation in whitened coordinates, can be corrected on the fly and were empirically observed to converge to the reference posterior.
More expressive Gaussian process (\GPabbr) surrogates on top of the Laplace approximation further tightened the rate approximation and improved sampling efficiency across the board.
While the Gaussian process model offered clear benefits, incorporating derivative observations and adaptive refinement yielded only modest additional gains in performance for our benchmark problem.

We concluded with a comparison of the \ZZS\ and the \BPS\ within the surrogate–assisted framework.
The \BPS\ had the upper hand regarding the posterior mean estimation error, the \ZZS\ attained better performance in terms of variance estimation, and both algorithms were largely on par regarding the Wasserstein distance.
The \BPS\ exhibited a higher effective sample size per model evaluation, particularly evident in higher dimensions.
Both \PDMP\ samplers outperformed a well tuned Random Walk Metropolis baseline and the popular No-U-Turn Sampler in all considered metrics, across all settings.

This study was restricted to smooth, unimodal posteriors arising from an elliptic PDE with an analytic forward map and moderate parameter dimensions.
A limitation of the employed Gaussian process surrogates is their cubic training cost in the number of design points, which limits their applicability in very high dimensions.
Moreover, while our empirical evidence supported convergence of the surrogate–assisted \PDMP\ samplers under the proposed correction scheme, a theoretical analysis of convergence for adaptive and non-strict upper bounds remains an open problem.

Future research directions include extending the approach to more realistic \PDE-governed inverse problems, higher-dimensional parameter spaces, and possibly multimodal posteriors.
On the surrogate side, sparse or low-rank Gaussian process constructions or neural surrogates could alleviate the cubic scaling while preserving global accuracy.
Finally, a rigorous theoretical treatment of convergence properties for the proposed surrogate–assisted \PDMP\ samplers would provide valuable insights and guidance for practical applications.
Nevertheless, this work establishes the first successful application of \PDMP\ sampling to \PDE-governed Bayesian inverse problems, demonstrating their viability for a wide range of scientific and engineering applications.
 
\bmhead{CRediT Authorship Contribution Statement}
L. Riccius: Conceptualization, Methodology, Software, Validation, Investigation, Visualization, Writing - Original Draft.
I.B.C.M. Rocha: Conceptualization, Methodology, Supervision, Writing - Review \& Editing.
J. Bierkens: Methodology, Writing - Review \& Editing.
H. Kekkonen: Writing - Review \& Editing.
F.P. van der Meer: Conceptualization, Methodology, Supervision, Writing - Review \& Editing.

\bmhead{Acknowledgements}
This work is supported by the TU Delft AI Labs programme through the SLIMM lab.

\bmhead{Code Availability}
The code used in this study is available at \href{github.com/SLIMM-Lab/pdmp.git}{\texttt{github.com/SLIMM-Lab/pdmp.git}}.

\begin{appendices}
\section{Influence of the Shrinkage Parameter \(\beta\)}\label{sec:influence_shrinkage}
\begin{figure}
    \centering
    \hspace{0.02\textwidth}\columnlabel{\(d=2\)}
    \columnlabel{\(d=5\)}
    \columnlabel{\(d=10\)}\\
    \rowlabel{\hspace{1.2cm} Constant}
    \subcaptionbox{\label{fig:shrink_const_2d}}[0.32\textwidth]{\includegraphics[width=\linewidth]{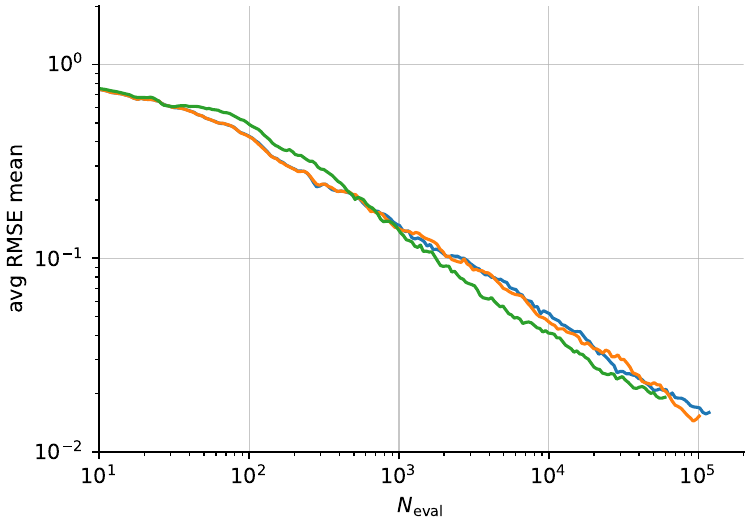}}\subcaptionbox{\label{fig:shrink_const_5d}}[0.32\textwidth]{\includegraphics[width=\linewidth]{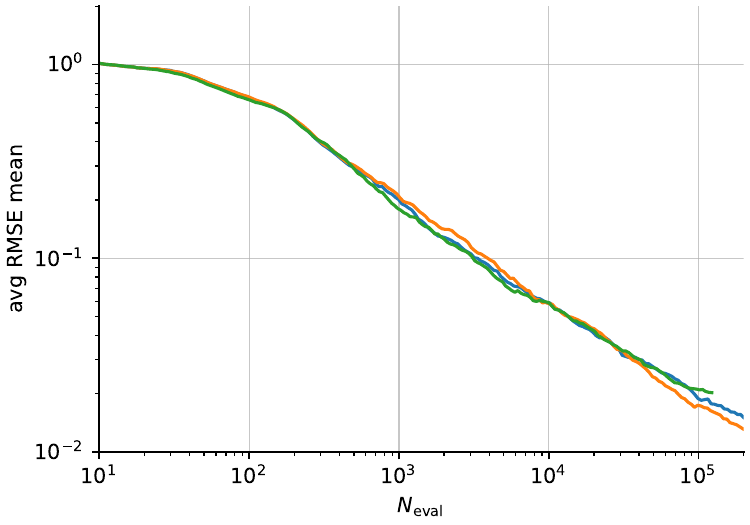}}\subcaptionbox{\label{fig:shrink_const_10d}}[0.32\textwidth]{\includegraphics[width=\linewidth]{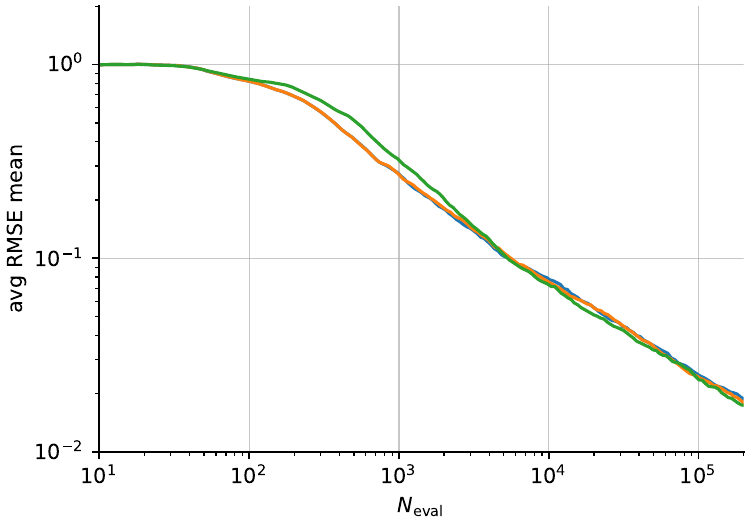}}\\
    \rowlabel{\hspace{1.2cm} Laplace}
    \subcaptionbox{\label{fig:shrink_lap_2d}}[0.32\textwidth]{\includegraphics[width=\linewidth]{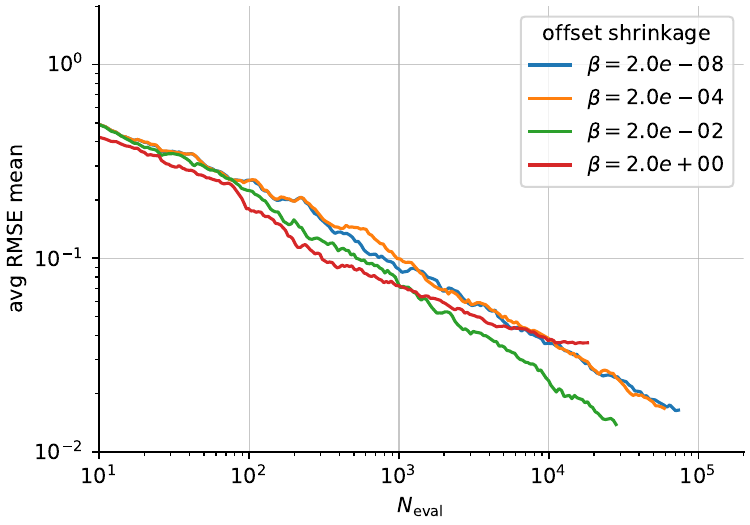}}\subcaptionbox{\label{fig:shrink_lap_5d}}[0.32\textwidth]{\includegraphics[width=\linewidth]{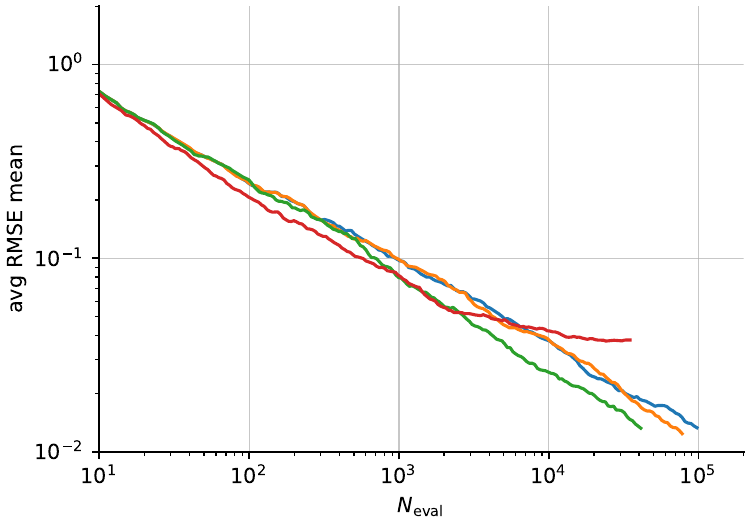}}\subcaptionbox{\label{fig:shrink_lap_10d}}[0.32\textwidth]{\includegraphics[width=\linewidth]{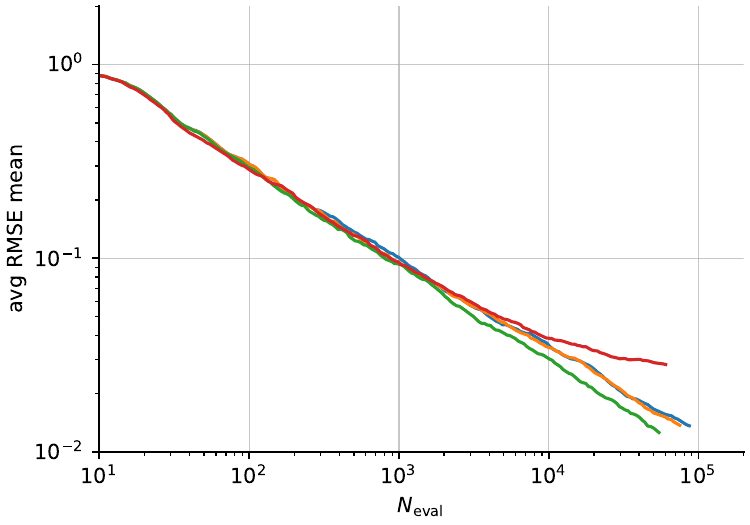}}\\
    \rowlabel{\hspace{1.2cm} \GPabbr\ 25}
    \subcaptionbox{\label{fig:shrink_gp_25_2d}}[0.32\textwidth]{\includegraphics[width=\linewidth]{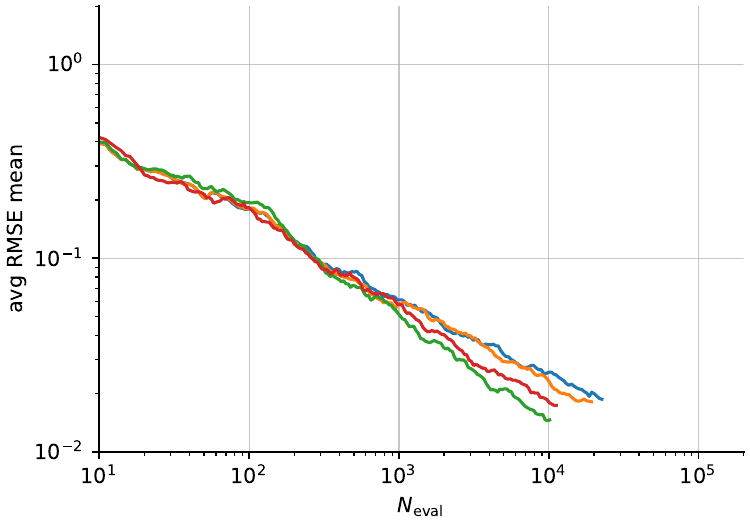}}\subcaptionbox{\label{fig:shrink_gp_25_5d}}[0.32\textwidth]{\includegraphics[width=\linewidth]{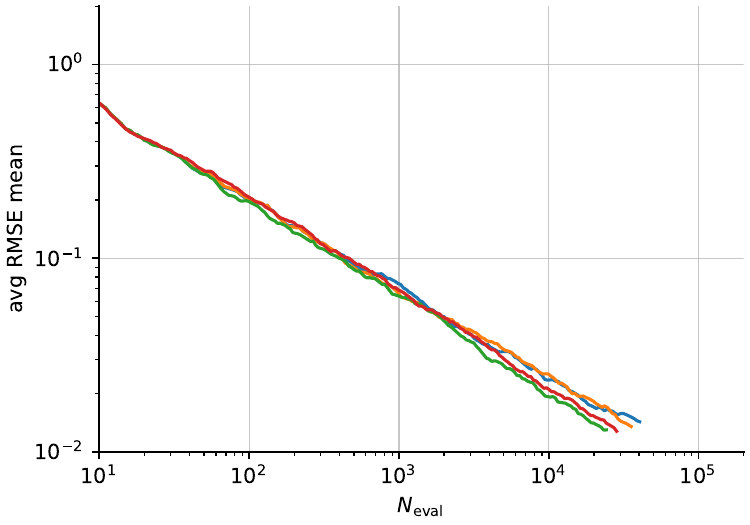}}\subcaptionbox{\label{fig:shrink_gp_25_10d}}[0.32\textwidth]{\includegraphics[width=\linewidth]{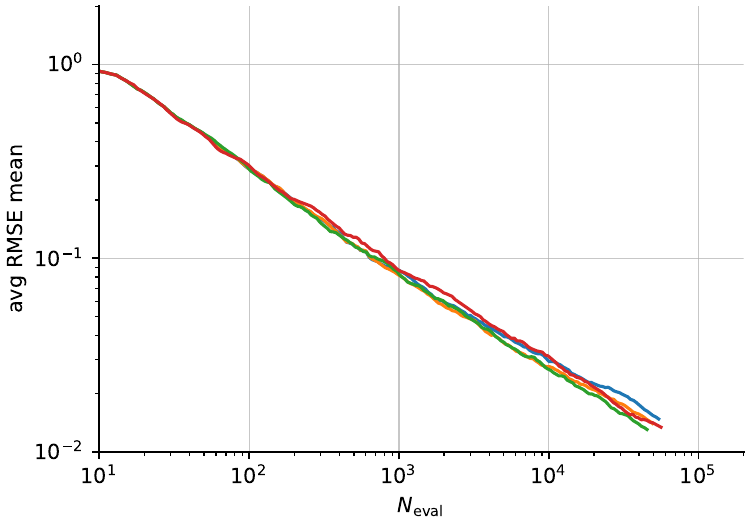}}\\
    \rowlabel{\hspace{1.2cm} \GPabbr\ 100}
    \subcaptionbox{\label{fig:shrink_gp_100_2d}}[0.32\textwidth]{\includegraphics[width=\linewidth]{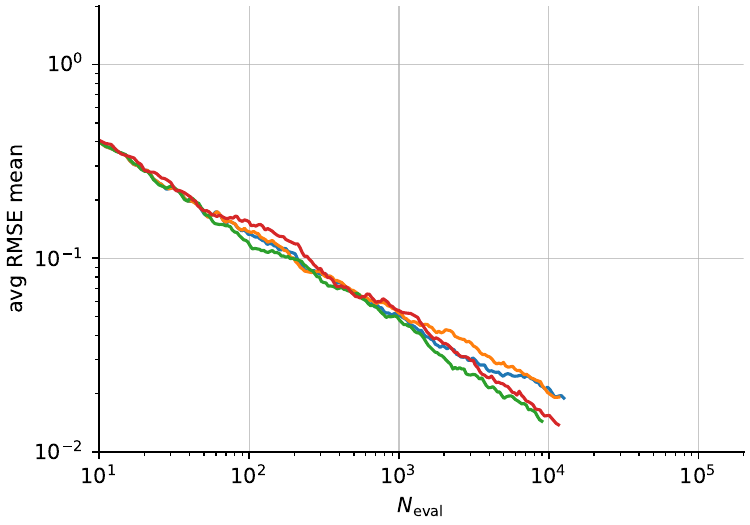}}\subcaptionbox{\label{fig:shrink_gp_100_5d}}[0.32\textwidth]{\includegraphics[width=\linewidth]{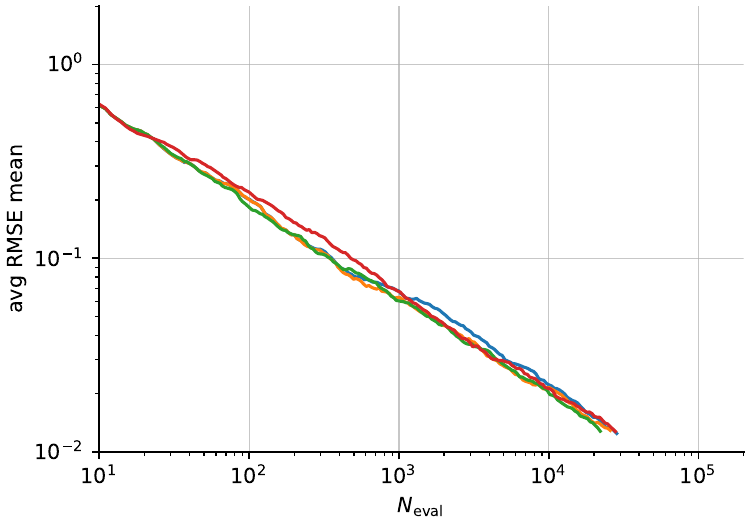}}\subcaptionbox{\label{fig:shrink_gp_100_10d}}[0.32\textwidth]{\includegraphics[width=\linewidth]{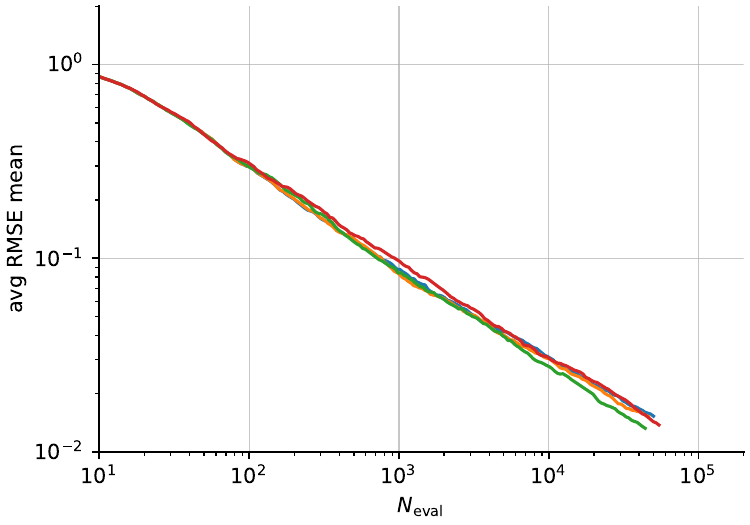}}\caption{
        Average \RMSE\ of the posterior mean estimate as a function of the number of model evaluations for different shrinkage parameters and surrogate models.
    }
    \label{fig:influence_shrinkage}
\end{figure}

For completeness, we investigate the influence of the shrinkage parameter \(\decay\) in \cref{eq:offset_decay} on the performance of the \ZZS\ with different surrogate models.
We include the constant surrogate model, a Laplace approximation, and two Gaussian process surrogate models with 25 and 100 training points per latent space dimension, respectively.
For each surrogate model—represented by a row in \cref{fig:influence_shrinkage}—we report the average \RMSE\ of the posterior mean estimate across a range of shrinkage parameter values \(\beta\).
The experiments are conducted in \(d=2\), \(d=5\), and \(d=10\) dimensions and shown as the columns of \cref{fig:influence_shrinkage}.

As a first observation, the \ZZS\ converges with all surrogate models when the shrinkage parameter is sufficiently small, i.e., \(\decay \leq \num{2d-2}\).
A second observation is that the \ZZS\ converges for all \(\decay\) values when using \GPabbr\ surrogate models, which are the most flexible in approximating the rate function.
For the Laplace model, the \ZZS\ fails to converge for \(\decay \geq \num{2}\).
In these cases, the \RMSE\ stagnates after a few hundred model evaluations, pointing to a persistent model bias.
The constant surrogate model performs worst.
It not only fails to converge for \(\decay \leq \num{2e-2}\) for all dimensionalities, but in fact diverges.
Results for \(\decay \geq \num{2e-2}\) are omitted, as the \PDMP\ simulation terminated prematurely for all random seeds due to numerical issues caused by large offset values.

\begin{figure}
    \centering
    \subcaptionbox{\label{fig:shrink_const_inspection_far}}[0.33\textwidth]{\includegraphics[width=\linewidth]{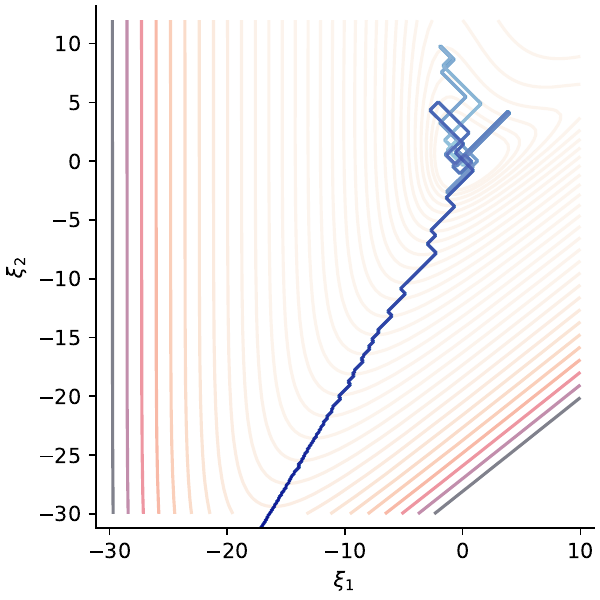}}\subcaptionbox{\label{fig:shrink_const_inspection_close}}[0.33\textwidth]{\includegraphics[width=\linewidth]{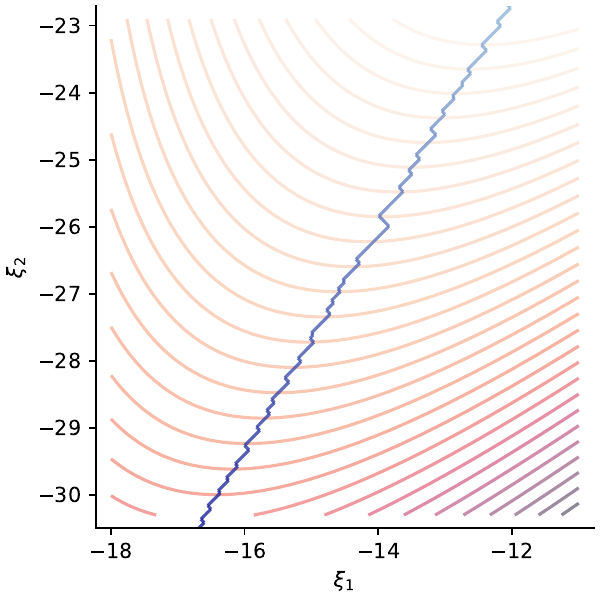}}\subcaptionbox{\label{fig:shrink_const_inspection_vel}}[0.33\textwidth]{\includegraphics[width=\linewidth]{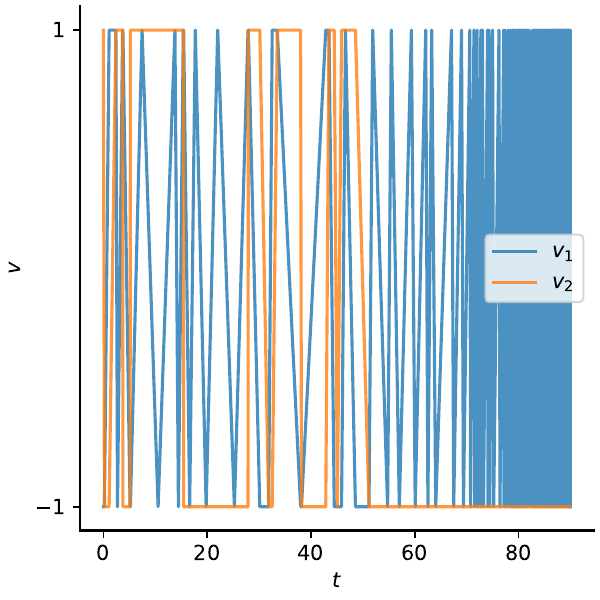}}\caption{
        Inspection of Zig-Zag sampler's path for the constant surrogate model with shrinkage parameter \(\decay = 2 \cdot 10^{-1}\).
        (a) Path and log-density.
        (b) Behaviour of the sampler in the tail.
        (c) Velocities of the sampler.
        The time is indicated by the colour gradient from light to dark in (a) and (b).
    }
    \label{fig:shrink_const_inspection}
\end{figure}

To better understand the failure of the constant surrogate with large shrinkage parameters, we inspect the trajectories of the \ZZS\@.
We consider a representative realization of the two-dimensional case with \(\decay = \num{2e-1}\).
After an initial phase of expected behaviour, during which the sampler explores the high-density region of the distribution, it suddenly drifts toward the lower-left corner of the domain, as shown in \cref{fig:shrink_const_inspection_far}.
The close-up in \cref{fig:shrink_const_inspection_close} reveals that the sampler oscillates around a line of points with zero gradient in the \(\apos_1\) direction, while steadily moving in the negative \(\apos_2\) direction.
\cref{fig:shrink_const_inspection_vel} illustrates the time interval just before and after this departure: both velocity components flip regularly until about \(t=\num{50}\), after which \(\vel_2\) remains fixed at \num{-1}, while \(\vel_1\) continues flipping at an increasing rate.

This behaviour can be traced to the offset dynamics.
The large offset in the \(\vel_1\) component drives frequent events, but the smaller offset in the \(\vel_2\) component quickly shrinks to zero for larger \(\decay\).
At the same time, the offset in the \(\vel_1\) component continues to grow, as we move further away from the high-density region.
As a result, the proposal mechanism fails to generate events in the \(\vel_2\) direction, causing the sampler to drift downward without correction.
Since the offset is only updated after an event in the corresponding direction and such an event never occurs, the sampler diverges.

When the shrinkage parameter is sufficiently small, the \PDMP\ spends an initial phase adjusting to an appropriate offset value, visible as a plateau at the start of the \RMSE\ curves in \cref{fig:shrink_const_2d,fig:shrink_const_5d,fig:shrink_const_10d}.
Once this value is established, it remains largely stable with only minor decreases over time, allowing the sampler to slowly converge.
For the more accurate and therefore more robust \GPabbr\ surrogates, runs with large values of \(\decay\) also converge.
In these cases, it appears beneficial that the offset recovers quickly after an unlikely event in the distribution’s tail, which forces a jump in the offset.

 \section{Bouncy Particle Sampler Refreshment Rate}\label{subsec:bps_refreshment_rate}
Before we can put the \BPS\ to the test, we need to determine a suitable refreshment rate \(\refreshrate\), which determines how often the velocity is resampled from its prior distribution.
A high refreshment rate leads to more frequent updates of this kind, which ensures ergodicity of the sampler.
However, it also makes the process more diffusive, more akin to a random walk, and thus less efficient.
The optimal refreshment rate is therefore a compromise between these two effects.
We investigate the influence of the refreshment rate on the performance of the \BPS\ with a \GPabbr\ surrogate model with \num{25} points per latent space dimension.
We consider refreshment rates \(\refreshrate \in \braces{\num{e-3},\num{e-2},\num{e-1},\num{e0}}\).

\begin{figure}
    \centering
    \subcaptionbox{\(d=2\) \label{fig:bps_refreshment_2d_mean}}[0.33\textwidth]{\includegraphics[width=\linewidth]{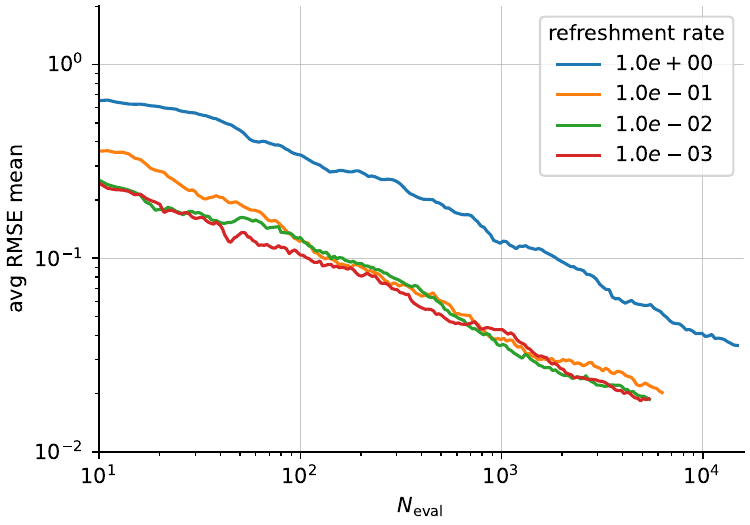}}\subcaptionbox{\(d=5\) \label{fig:bps_refreshment_5d_mean}}[0.33\textwidth]{\includegraphics[width=\linewidth]{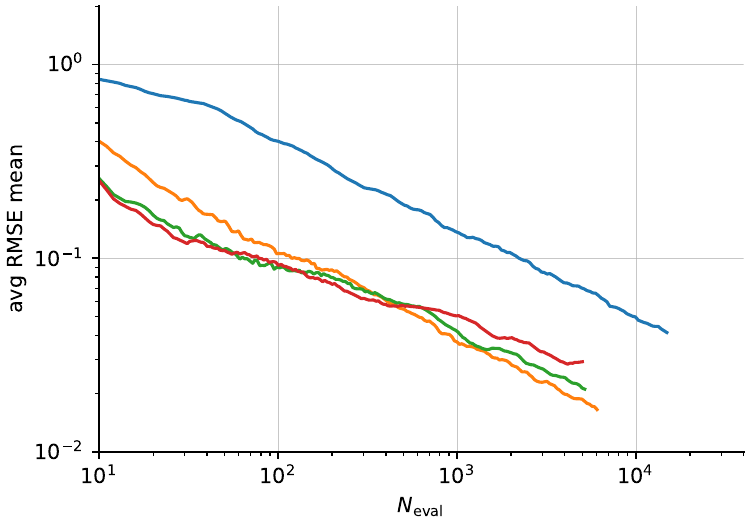}}\subcaptionbox{\(d=10\) \label{fig:bps_refreshment_10d_mean}}[0.33\textwidth]{\includegraphics[width=\linewidth]{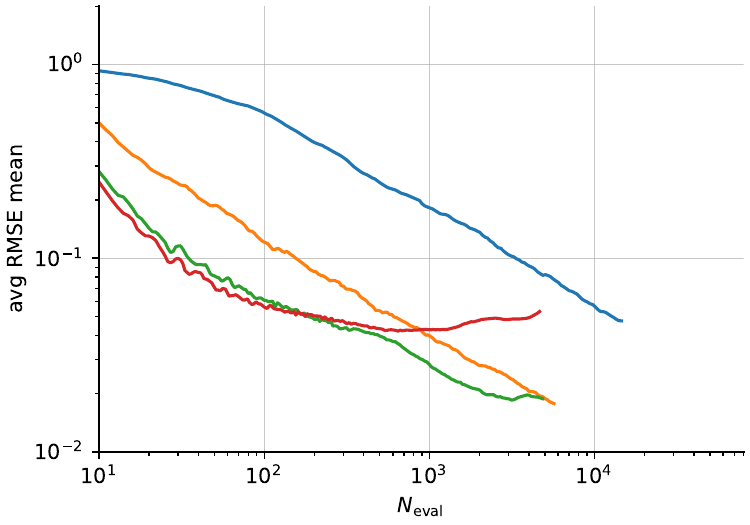}}\\
    \subcaptionbox{\(d=2\) \label{fig:bps_refreshment_2d_var}}[0.33\textwidth]{\includegraphics[width=\linewidth]{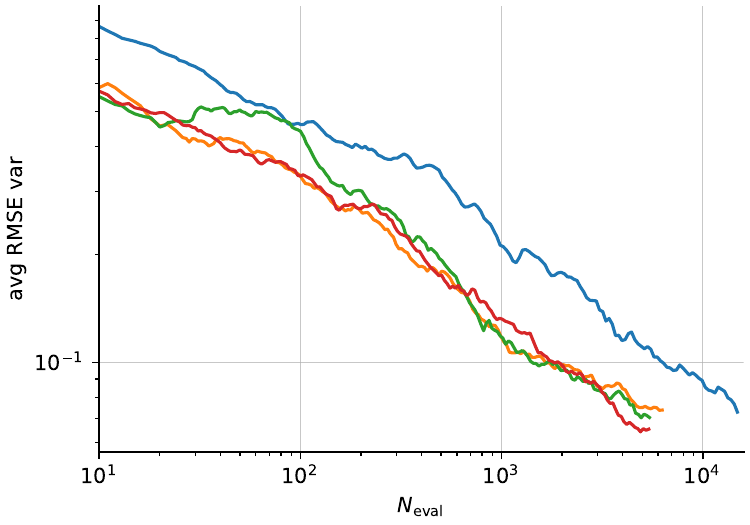}}\subcaptionbox{\(d=5\) \label{fig:bps_refreshment_5d_var}}[0.33\textwidth]{\includegraphics[width=\linewidth]{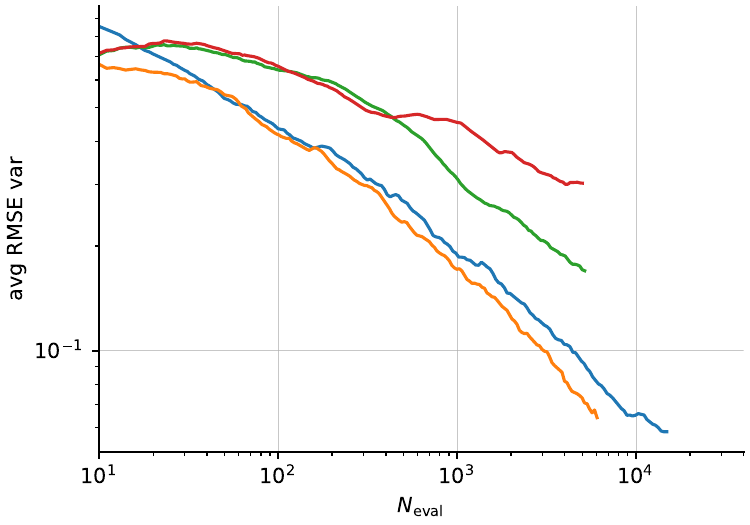}}\subcaptionbox{\(d=10\) \label{fig:bps_refreshment_10d_var}}[0.33\textwidth]{\includegraphics[width=\linewidth]{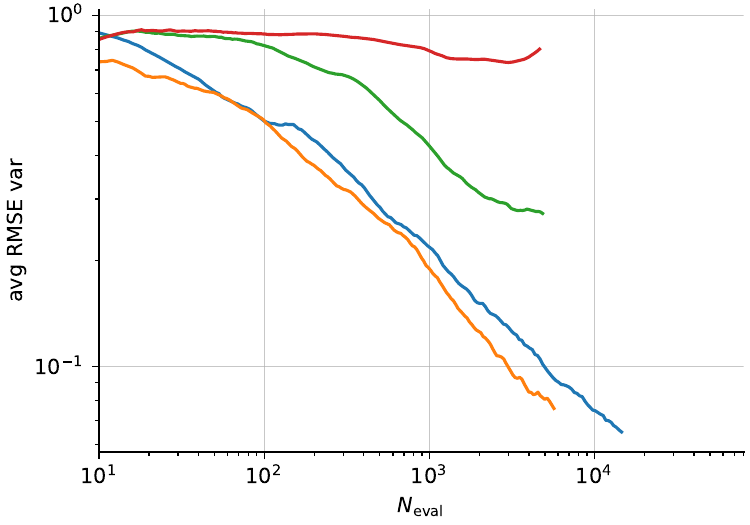}}\caption{
        Comparison of different \BPS\ refreshment rates in terms of \RMSE\ of the mean (top row) and variance (bottom row).
        Moderate refreshment rates yield the best performance across both metrics.
    }
    \label{fig:bps_refreshment}
\end{figure}

The results in \cref{fig:bps_refreshment} show that the refreshment rate has a noticeable influence on the performance of the \BPS\@.
High refreshment rates lead to slower convergence in terms of the posterior mean estimate in the top row of \cref{fig:bps_refreshment}.
The performance in terms of accuracy of the mean is relatively stable for \(\refreshrate \leq \num{e-1}\), suggesting diminishing returns for lower refreshment rates.
A look at the variance in the bottom row of \cref{fig:bps_refreshment}, however, reveals a different picture:
very low refreshment rates lead to poor approximation of the posterior variance, suggesting the ergodicity of the sampler is compromised in the absence of refreshments.
We see that setting the refreshment rate too high leads to slow exploration, and choosing it too low to poor coverage, but there seems to be a sweet spot around \(\refreshrate = \num{e-1}\).
 
\end{appendices}
\FloatBarrier

\bibliography{export}

\end{document}